\documentclass[twocolumn,tighten]{aastex63}
\usepackage{natbib}
\usepackage{epsfig}
\usepackage{amsmath}
\usepackage{xcolor}
\usepackage{hyperref}
\usepackage{booktabs}
\usepackage{xtab}
\usepackage{afterpage}
\usepackage[encapsulated]{CJK}
\usepackage{ucs}
\usepackage[utf8x]{inputenc}
\usepackage{url}
\usepackage{lineno}
\usepackage[caption=false]{subfig}

\begin{document}

\newcommand{\cotwo}{\mbox{\rm CO($2\text{--}1$)}} 

\newcommand{\halpha}{\mbox{\rm H$\alpha$}}

\newcommand{\xco}{\mbox{$X_{\rm CO}$}}

\newcommand{\xcounits}{\mbox{cm$^{-2}$ (K~km~s$^{-1}$)$^{-1}$}}

\newcommand{\acounits}{\mbox{M$_\odot$~pc$^{-2}$ (K~km~s$^{-1}$)$^{-1}$}}

\newcommand{\sigsfr}{\mbox{$\Sigma_{\rm SFR}$}}

\newcommand{\ANU}{\affiliation{Research School of Astronomy and Astrophysics, Australian National University, Canberra, ACT 2611, Australia}}

\newcommand{\ARI}{\affil{Astronomisches Rechen-Institut, Zentrum f\"{u}r Astronomie der Universit\"{a}t Heidelberg, M\"{o}nchhofstra\ss e 12-14, D-69120 Heidelberg, Germany}}

\newcommand{\ASTROThreeD}{\affiliation{ARC Centre of Excellence for All Sky Astrophysics in 3 Dimensions (ASTRO 3D), Australia}}

\newcommand{\Bonn}{\affiliation{Argelander-Institut f\"{u}r Astronomie, Universit\"{a}t Bonn, Auf dem H\"{u}gel 71, 53121 Bonn, Germany}}

\newcommand{\Carnegie}{\affiliation{Observatories of the Carnegie Institution for Science, 813 Santa Barbara Street, Pasadena, CA 91101, USA}}

\newcommand{\CCAPP}{\affil{Center for Cosmology and Astroparticle Physics, 191 West Woodruff Avenue, Columbus, OH 43210, USA}}

\newcommand{\CITA}{\affiliation{Canadian Institute for Theoretical Astrophysics (CITA), University of Toronto, 60 St George Street, Toronto, ON M5S 3H8, Canada}}

\newcommand{\CNRS}{\affil{CNRS, IRAP, 9 Av. du Colonel Roche, BP 44346, F-31028 Toulouse cedex 4, France}}

\newcommand{\COOL}{\affil{Cosmic Origins Of Life (COOL) Research DAO, coolresearch.io}}

\newcommand{\ITA}{\affil{Instit\"ut f\"{u}r Theoretische Astrophysik, Zentrum f\"{u}r Astronomie der Universit\"{a}t Heidelberg, Albert-Ueberle-Strasse 2, 69120 Heidelberg, Germany}}

\newcommand{\McMaster}{\affiliation{Department of Physics and Astronomy, McMaster University, 1280 Main Street West, Hamilton, ON L8S 4M1, Canada}}

\newcommand{\MPIA}{\affil{Max Planck Institut f\"{u}r Astronomie, K\"{o}nigstuhl 17, 69117 Heidelberg, Germany}}

\newcommand{\NRAO}{\affil{National Radio Astronomy Observatory, 520 Edgemont Road, Charlottesville, VA 22903-2475, USA}}

\newcommand{\OSU}{\affil{Department of Astronomy, The Ohio State University, 140 West 18th Avenue, Columbus, Ohio 43210, USA}}

\newcommand{\OSUPhysics}{\affil{Department of Physics, 191 West Woodruff Avenue, Columbus, OH 43210, USA}}

\newcommand{\Toulouse}{\affil{Universit\'{e} de Toulouse, UPS-OMP, IRAP, F-31028 Toulouse cedex 4, France}}

\newcommand{\UChile}{\affiliation{Departamento de Astronom\'{i}a, Universidad de Chile, Camino del Observatorio 1515, Las Condes, Santiago, Chile}}

\newcommand{\uwyo}{\affil{Department of Physics \& Astronomy, University of Wyoming, Laramie, WY 82071 USA}}

\shorttitle{Comparing the locations of supernovae to CO~(2-1) emission in their host galaxies}
\shortauthors{Mayker Chen et al.}


\title{Comparing the Locations of Supernovae to CO~(2-1) Emission in their Host Galaxies}

\begin{abstract}
We measure the molecular gas environment near recent ($< 100$ yr old) supernovae (SNe) using $\sim1''$ or $\leq 150$~pc resolution CO~(2-1) maps from the PHANGS-ALMA survey of nearby star-forming galaxies. This is arguably the first such study to approach the scales of individual massive molecular clouds ($M_{\rm mol} \gtrsim 10^{5.3}$ M$_{\odot}$). Using the Open Supernova Catalog (OSC), we identify 63 SNe within the PHANGS--ALMA footprint. We detect CO~(2-1) emission near $\sim60\%$ of the sample at 150~pc resolution, compared to $\sim35\%$ of map pixels with CO~(2-1) emission,  and up to $\sim95\%$ of the SNe at 1~kpc resolution compared to $\sim80\%$ of map pixels with CO~(2-1) emission. We expect the $\sim60\%$ of SNe within the same 150~pc beam as a GMC will likely interact with these clouds in the future, consistent with the observation of widespread SN-molecular gas interaction in the Milky Way, while the other $\sim40\%$ of SNe without strong CO~(2-1) detections will deposit their energy in the diffuse interstellar medium (ISM), perhaps helping drive large-scale turbulence or galactic outflows. Broken down by type, we detect CO~(2-1) emission at the sites of $\sim85\%$ of our 9 stripped-envelope SNe (SESNe), $\sim40\%$ of our 34 Type II SNe, and $\sim35\%$ of our 13 Type Ia SNe, indicating that SESNe are most closely associated with the brightest CO~(2-1) emitting regions in our sample. Our results confirm that SN explosions are not restricted to only the densest gas, and instead exert feedback across a wide range of molecular gas densities.

\end{abstract}


\correspondingauthor{Ness Mayker Chen}
\email{maykerchen.1@osu.edu}

\author[0000-0002-5993-6685]{Ness~Mayker~Chen}
\OSU \CCAPP

\author[0000-0002-2545-1700]{Adam~K.~Leroy}
\OSU\CCAPP

\author[0000-0002-1790-3148]{Laura A. Lopez}
\OSU \CCAPP

\author[0000-0003-4826-9079]{Samantha Benincasa}
\OSU \CCAPP

\author[0000-0002-5635-5180]{M\'{e}lanie~Chevance}
\ITA
\COOL

\author[0000-0001-6708-1317]{Simon~C.~O.~Glover}
\ITA

\author[0000-0002-9181-1161]{Annie~Hughes}
\CNRS
\Toulouse

\author[0000-0001-6551-3091]{Kathryn~Kreckel}
\ARI

\author[0000-0002-4781-7291]{Sumit Sarbadhicary}
\OSU \CCAPP

\author[0000-0003-0378-4667]{Jiayi~Sun \begin{CJK*}{UTF8}{gbsn}(孙嘉懿)\end{CJK*}}
\altaffiliation{CITA National Fellow}
\OSU
\McMaster
\CITA

\author[0000-0003-2377-9574]{Todd A. Thompson}
\OSU \CCAPP \OSUPhysics

\author[0000-0003-4161-2639]{Dyas~Utomo}
\NRAO

\author{Frank Bigiel}
\Bonn

\author[0000-0003-4218-3944]{Guillermo A. Blanc}
\Carnegie
\UChile

\author[0000-0002-5782-9093]{Daniel~A.~Dale}
\uwyo

\author[0000-0002-3247-5321]{Kathryn Grasha}
\altaffiliation{ARC DECRA Fellow}
\ANU
\ASTROThreeD

\author[0000-0002-8804-0212]{J.~M.~Diederik~Kruijssen}
\COOL

\author[0000-0002-1370-6964]{Hsi-An Pan}
\affiliation{Department of Physics, Tamkang University, No.151, Yingzhuan Road, Tamsui District, New Taipei City 251301, Taiwan}

\author{Miguel Querejeta}

\author[0000-0002-3933-7677]{Eva Schinnerer}
\MPIA

\author[0000-0002-7365-5791]{Elizabeth~J. Watkins}
\ARI

\author[0000-0002-0012-2142]{Thomas G. Williams}\MPIA


\section{Introduction}
\label{sec:intro}

\subsection{Overview}
Feedback from supernovae (SNe) is among the most influential processes in galaxy evolution. SNe launch galactic winds \citep[e.g.,][]{Chevalier1985,Veilleux2020}, heat the interstellar medium \cite[ISM, e.g.,][]{McKee1977}, regulate star formation \citep[e.g.,][]{Kim2015}, and help set the balance between the gravitational and pressure forces within the disks of galaxies \citep[e.g.,][]{Thompson2005,Ostriker2011,Sun2020}. However, the exact effects of SNe on galaxies and giant molecular clouds (GMCs) depend on the locations where the SNe occur. SNe that occur in environments full of low-density gas cool slowly and hence are efficient at heating and ejecting material out of the galaxy, resulting in hot, diffuse gas outflows \citep[e.g.][]{Martizzi2016, Andersson2020, Lopez2020}.  Conversely, SNe that occur in or near dense clouds can destroy GMCs and shut down star formation in these areas but cool more rapidly and hence are less effective at driving outflows \citep{Walch2015,Keller2020,Lu2020}. As a result, the impact of SNe on the ISM and their parent galaxies depends on the relative spatial configuration of the SNe, the gas, and the density distribution of that gas. However this relative configuration is not immediately obvious from first principles. Both thermonuclear (or Type Ia) SNe and core-collapse SNe (CCSNe) explode with delay times -- i.e., the time since the formation of the stellar population -- that can be long relative to the timescales over which the ISM evolves (see \S \ref{sec:background}). Further complicating this picture is the finding that stars often form in clusters \citep{Kruijssen2012,Krumholz2019,Adamo2020} and theoretical models show that the clustering of SNe and feedback leads to more effective driving of outflows \citep{Keller2022, Orr2022}.

Despite the importance of this topic, we do not yet have a clear observational picture of where SNe explode relative to the gas in their host galaxies, partly because sensitive, high-resolution observation of cold, star-forming interstellar gas in distant galaxies was not possible until the advent of the Atacama Large Millimeter/Submillimeter Array (ALMA). Previous studies of SN environments have carried out detailed analyses focused on the stellar populations at the sites of SNe using ground-based optical telescopes \citep{Anderson2008,Anderson2009,Anderson2012}. In the more spatially-resolved environment of the Milky Way and Local Group galaxies, studies have focused on the interactions of supernova remnants (SNRs) and gas clouds \citep[e.g.,][]{Jiang2010}. There have also been studies of the properties and distributions of H{\sc ii} regions and clouds, which suggest that pre-SN feedback mechanisms, such as stellar winds and ionizing radiation, can exert a strong influence \citep[e.g.,][]{Lopez2011, Lopez2014, Kruijssen2019, Chevance2020, Barnes2021, McLeod2021, Olivier2021, Chevance2022}. In the Milky Way, individual, young, massive star clusters show evidence of substantial feedback but no SN explosions \citep[e.g.,][]{Zeidler2016, Beasor2021}. 

These studies only provide indirect or incomplete information on the gas in the environments where stars actually explode, or in the case of SNRs, measurements that occur after the SNe have processed the ISM \citep[e.g.,][]{Sarbadhicary2017}. A more direct, complementary measurement would be to measure the gas content at the sites of very recent SN explosions, taking advantage of high-resolution spectral-line observations enabled by ALMA. However, there has been only a small amount of such work and most of that has used low-resolution ($\sim$kpc scale) gas data. 

For example, in one of the only direct measurements of gas at the sites of SNe to date, \citet{Galbany2017} measured CO~(2-1) emission from the sites of 26 SNe in the EDGE-CALIFA survey \citep{Bolatto2017}. While pioneering, this work was limited to a median resolution of $\sim1.4$~kpc. This is far larger than the spatial scales on which SN feedback occurs ($\sim100-200$~pc) and is not sufficient to localize the relative positions of individual SNe and GMCs. 

A key next step is to measure the molecular gas present in the local environments of SN explosion sites on GMC scales ($\sim10-100$~pc). Such measurements will more directly access the mean densities where SNe explode and will help inform SN placement in galaxy evolution simulations. By sorting these measurements by SN type, one can study how the explosion landscape depends on the lifetime of the SN progenitor and how each class of SNe impacts its galactic environment. Given observations with high enough resolution, one might even directly measure the fraction of SNe that may help to disperse GMCs. Such measurements will be critical to assess the quantitative role of SNe in the evolution of the ISM, the launching of galactic winds, and of setting star formation rates (SFRs) and efficiencies. The obstacle has been assembling a large enough sample of recent SNe with accompanying high-resolution gas observations. 

In this paper, we measure the molecular gas content at high ($\leq$150~pc) spatial resolution at the sites of 63 recent ($\leq 100$\,yr) SN explosions. To do this, we take advantage of the PHANGS--ALMA\footnote{``Physics at High Angular resolution in Nearby GalaxieS with ALMA". For more information, see www.phangs.org.} CO~(2-1) survey \citep{Leroy2021}, which mapped the molecular gas content of 90 of the nearest galaxies at $50-150$~pc resolution. 

\label{sec:background}
\subsection{Background}

This work focuses on characterizing the local environments of recent SNe. Previous SN environment studies \citep{James2006, Anderson2008, Anderson2009, Kelly2008, Habergham2010, Habergham2012, Galbany2012, Galbany2014, Galbany2017, Anderson2012, Kangas2013, Anderson2015a, Anderson2015b, AudcentRoss2020, Cronin2021} have employed pixel statistics and radial distribution comparisons to cross-correlate different types of SNe with tracers of star formation and stellar populations and assess the impact and likely progenitors of SNe. See \cite{Anderson2015b} for a more thorough review.
 
These studies have helped inform our emerging picture of SNe. Core-collapse supernovae (CCSNe, such as Type Ib/Ic and Type II) are triggered by the collapse of the inert Fe-core of massive stars, occurring on timescales that decrease as the progenitor mass increases, down to a minimum of 4 Myr for stars of 60 $M_\odot$ and above (see reviews by \citealt{Smartt2009, Muller2016}).  Stripped-envelope SNe (SESNe), such as SNe IIb, Ib, and Ic, are CCSNe that have lost some or all of their hydrogen (SNe IIb \& SNe Ib) and helium (SNe Ic) envelopes prior to explosion. SESNe have high-mass progenitors, with observations that support an increasing mass range from SNe IIb-Ib-Ic, with SNe Ic coming from the youngest population of progenitors \citep{Kelly2008, Anderson2009}. Because they arise from these very young stars, SNe Ic have the strongest association with other signatures of recent star formation out of all of the SNe types, and SESNe have stronger associations with star-forming regions than SNe II \citep{Anderson2012, Galbany2017}.

SNe Ia are the thermonuclear explosions of white dwarf stars in binary systems (see review by \citealt{Maoz2014}). These explosions have a comparatively long delay time relative to CCSNe, often described as a power law distribution with approximately equal SNe per logarithmic time interval \citep{Maoz2014, Maoz2017}. This implies that many SNe Ia will explode long after their natal cloud and stellar population have dissipated. Due to the longer delay time of SNe Ia, their explosions should track the overall distribution of stellar mass better than recent star formation, reflecting that they arise from stars with a broad range of ages \citep[e.g., see][]{Anderson2015a}. This may also lead us to expect a relatively weak correlation between the location of SNe Ia and the gas clouds that host star formation.

The massive stars that produce CCSNe are formed in GMCs. These stars have lifetimes ranging from $\sim$ 3--50~Myr\footnote{Calculations from \cite{Leitherer2014} find that the approximate time from zero-age main sequence to the first SN explosion is $\sim 4$~Myr. The shortest lifetimes of the most massive stars converge to $\sim$ 2.5 Myr \citep{Choi2017}, but stars more massive than 30 $M_\odot$ have decreased explodability \citep{Sukhbold2016}. It seems likely that the first SN explosion is $\sim 3{-}5$ Myr after the birth of a fully sampled initial mass function (IMF), but in some regions this may be longer due to incomplete sampling of the IMF \citep{Chevance2022}.}, with some rarer instances extending the tail of the delay time distribution past 100~Myr\footnote{These include electron-capture supernovae arising from super-AGB stars \citep{Prieto2008, Thompson2009} and the merger of $4M_\odot + 4M_\odot$ binaries, which produce a tail of explosions in  binary population synthesis out to 100-200~Myr \citep{Zapartas2017}}. Because of the short lifespans of the higher-mass range of CCSNe progenitors, one would expect that a fraction of CCSNe should be closely associated with molecular gas. However there are two known complications that we must consider: pre-SN feedback on the molecular gas and the non-trivial fraction of these massive stars that have moved away from their birth site. 

Previous observational work has shown that massive stars exert strong feedback on their immediate environment in the form of winds and radiation even before they explode. This pre-SN feedback has been characterized by comparing individual pressures within \textsc{Hii} regions \citep[e.g.][]{Lopez2011, Pellegrini2011, Lopez2014, McLeod2020, Barnes2021, Olivier2021} and by contrasting the distributions and fluxes of H$\alpha$ to CO~(2-1) emission within star-forming regions of galaxies \citep{Schruba2010, Kruijssen2019, Schinnerer2019, Chevance2020, Chevance2022, Kim2022, Pan2022}. As discussed by \cite{Kruijssen2019} and \cite{Chevance2020,Chevance2022}, the existence of widespread H$\alpha$ emission without associated CO~(2-1) implies rapid clearing of molecular gas by stellar feedback. As a consequence of this, GMCs have been observed to have lifespans on the order of 5-30 Myr \citep[e.g., see][J. Turner et al. ApJ in review]{Kawamura2009, Fukui2010, Meidt2015, Corbelli2017, Grasha2018, Grasha2019, Kruijssen2019, Chevance2020, Kim2021, Kim2022}.

Previous theoretical work \citep[e.g.][]{Krumholz2009, Murray2010, Stinson2013, Geen2016, Gatto2017, Rahner2019, Lucas2020, Keller2022} has also emphasized the crucial role played by pre-SN feedback processes in regulating the star formation efficiency of GMCs, and the impact that the SN environment has on the effectiveness of SN feedback \citep[see e.g.][]{Walch2015, Iffrig2015, Ohlin2019, Andersson2020}. Modern simulations \citep[e.g.][]{Dobbs2013, Kim2018, Benincasa2020} that model stellar feedback find the majority of these clouds live only 5--7 Myr with only a small percentage surviving to 20 Myr. 

Even without clearing the parent cloud by pre-SN feedback, it may be possible for massive stars to separate from their birth site before they explode as SNe. Only the most massive O stars have lifetimes as short as 2.5 Myr, but most CCSNe have B star progenitors which have delay times on the order of 10-40 Myr. Assuming the relative velocity between progenitors and their natal molecular clouds (MCs) is on the order of the velocity dispersion of individual stellar associations (e.g. 10 km/s $\sim$ 10 pc/Myr), we would expect that they would move 100-400~pc before exploding in that delay time. Most of these stars will have traveled far enough away from their natal cloud to explode in diffuse regions of the galaxy. \cite{Aadland2018} found that in Local Group galaxies, red supergiants (progenitors of Type II SNe) are offset by nearly 200-600~pc from the nearest blue, main-sequence star, and twice as isolated as higher-mass stars (e.g., Wolf-Rayets). Previous observational work has shown that 20-30\% of OB stars appear in the field \citep{Oey2004, Gies1987} in regions that are not associated with GMCs. Up to 46\% of O stars and 4\% of B stars have been observed to have “runaway” velocities of at least 30 $km s^{-1}$, leading them to be found at large distances from their galactic birth site \citep{Stone1991, Russeil2017}. Therefore, in the scenario where GMCs are not destroyed by pre-SN feedback, there can still be a significant fraction of CCSNe that are not coincident with molecular gas because the stars have moved away.

Because of the complexities discussed above, a priori predictions of where SNe occur relative to the gas in galaxies are complicated. The uncertainty and new information about pre-SN feedback, motions of stars, progenitors, etc., adds layers of complexity to this problem. In this paper, we make a series of measurements aimed at improving our knowledge of where SNe explode via direct observations. We measure the surface density and mass of molecular gas at the sites of recent explosions, measure the distance to the nearest detectable concentration of molecular gas, and examine how these results depend on physical resolution. 

In \S \ref{sec:Methods}, we identify all SNe that have recently occurred in the footprint of the PHANGS--ALMA CO~(2-1) maps. In \S \ref{sec:Results}, we present our results as follows: In \S \ref{sec:detectionallres}, we measure the signal-to-noise of the CO~(2-1) emission at the location of each SN at multiple resolutions from 60~pc-1~kpc. This yields a basic measurement of how often molecular gas appears associated with the SN explosion sites and how this changes as a function of scale. In \S \ref{sec:detection150pc}, we compare the distribution of CO~(2-1) emission intensities at the location of the SN sample to the full intensity distribution in the galaxies at our most complete resolution of 150~pc. We estimate the corresponding mass and surface density of molecular gas. We then investigate how these distributions change as a function of SN type. In \S \ref{sec:detectionFiner},
we compare the molecular gas surface density distributions for the SN explosion sites and for the full set of pixels in the CO~(2-1) maps at resolutions from 60-150~pc. In \S \ref{sec:zooms} we look at $500 \times 500~pc$ cut-outs of each individual SN site at the highest resolution CO~(2-1) map available to see if the fraction of SN environments coincident with molecular gas changes at smaller scales and to classify the potential feedback from each SN. In \S \ref{sec:MCdistance}, we measure the distribution of distances from our SN sample to the nearest location with well-detected molecular gas and the distribution of molecular gas surface densities. We then compare these distributions to several model populations of SNe. Finally, in \S \ref{sec:Discussion}, we discuss the significance of our results.


\section{Methods}
\label{sec:Methods}

To explore the relationship between SNe and their natal GMCs, we first aim to identify all recent SNe that have exploded in the PHANGS--ALMA footprint and measure the molecular gas content of the ISM near these explosions. To do this, we select SNe from the Open Supernova Catalog (OSC; \citealt{Guillochon2017}), accessed on 2021 May 25, which includes almost all known recent SNe by blending a heterogeneous collection of all available previous surveys. 

Our initial sample consists of all of the OSC objects located within the PHANGS--ALMA CO~(2-1) mapping footprint that include a discovery date. By requiring a discovery date, we limit the sample mostly to SNe that have occurred in the last $\sim 20$ years. The earliest SN in our sample was recorded in 1901, and the median discovery date was 2003. PHANGS--ALMA has a resolution of $\sim 100$ pc, and even with high shock velocities of $\sim 5,000$~km~s$^{-1}$, a SN will only impact $\lesssim 1$~pc in 100 years. By contrast, if we included SNRs, which can remain detectable for $\sim 10^{5}$ years \citep{Sarbadhicary2017}, we would expect to see frequent signatures of interaction between SNRs and clouds and note that some SNe might have already cleared their parent cloud. We focus on recent SNe because we want to study the environments before the explosions have had time to strongly influence their surroundings. 

In Table \ref{tab:OSCSample}, we list the OSC object name, reported type, parent galaxy, and coordinates. We also include the reference paper for the astrometry and SN type or the discovery announcement if a reference paper with typing is not available\footnote{Note that for two of the OSC objects (PTSS-19clju and PSN+204.332083-29.896833) we are unable to find a reference, and those are not included in our final, working sample.}. Finally, we note whether the object is included in our working SN sample.

We caution that the OSC is heterogeneous. It includes all discovered SNe but without assurances of completeness. Many of the SNe in the OSC are discovered via targeted galaxy searches, which introduces a bias towards more massive host galaxies. Because more massive host galaxies have higher star formation rates and higher molecular gas surface densities, this bias could increase the frequency in which we find SNe coincident with molecular gas. SN searches, especially early searches, also show some bias against finding SNe near the bright center of galaxies \citep[e.g., see][]{Holoien2017}, and we would expect them to also be biased against finding the most heavily embedded SNe due to extinction. Because of this, we might miss SN populations located in high-extinction regions, which would decrease the fraction of SNe found coincident with molecular gas and limit the association that we measure between SNe and the highest density environments.

We consider galaxies targeted by the PHANGS--ALMA survey \citep{Leroy2021} which includes a representative sample of 90 relatively massive, actively star-forming, local ($d \leq 23$~Mpc) galaxies. This survey used the Atacama Large Millimeter/submillimeter Array (ALMA) to map CO~(2-1) emission on $1\arcsec\ \sim 100$~pc scales. CO~(2-1) emission is a standard tracer of the distribution of molecular gas in galaxies \citep[for a review see][]{Bolatto2013}, tracing the cold, dense, star-forming phase of the ISM. The PHANGS--ALMA survey covers $\sim70\%$ of the CO~(2-1) emission and active star formation, as traced by WISE emission, in each galaxy. We checked each SN entry in the OSC to see if the host was a member of PHANGS--ALMA or if the coordinates lay within the astrometric footprint of the PHANGS--ALMA CO~(2-1) integrated intensity map. We verified the host galaxy assignment and SN type classification via discovery/follow-up papers for each individual SN (see Table \ref{tab:OSCSample}). We then verified by eye that the SNe indeed lie within the CO~(2-1) maps. The PHANGS–ALMA footprint does not cover the outer disk regions of the galaxies surveyed. Because of this, we will miss any SNe that occur in these outer regions. This introduces a bias where fewer SNe will be associated with these lower-density environments. The PHANGS–ALMA survey also cannot “see” all of the molecular gas in the regions that are surveyed. The $\sim70\%$ completeness will cause a underreporting of SN-CO interactions occurring in low-density environments.

We find 78 OSC objects in 33 galaxies within the PHANGS--ALMA footprint (see Table \ref{tab:OSCSample}). Of these, we discard a luminous red nova (LRN) reported in the catalogue (AT2020nqq), 7 SN candidates (AT2018eoq, AT2019pck, AT2019npi, AT2019npd, AT2020cwh, AT2020hol, AT2020juh), the second entry of a SN reported by two groups under two different names (SN2019ehk -- classified as SN type Ib \& PTSS-19clju -- listed as unclassified), and two Type II SNe that occurred on the edge of the mapped region of NGC0628 (SN2003gd) and NGC4945 (SN2005af) where the CO~(2-1) data appeared noisier and the image quality poorer. (SN1992eu also falls along the edge of NGC1097's map, but we keep it in the sample because the map quality appeared fine at its position.) We also discard PSN$+$204.332083$-$29.896833 in NGC~5236, which has minimal reported data and is outside of the mapped region of the galaxy. Finally, we remove three SNe (SN1997bs, SN2019krl, SN2019qyl) from our sample because of uncertainty in their type classification and the liklihood that they are instead non-terminal explosions \citep{Adams2015, Andrews2021, Jencson2021, Ransome2021}. This brings our sample size to 63 SNe within 31 galaxies.

We are interested in the coincidence of SNe and molecular gas; this measurement will be limited by the resolution of the ALMA data and the positional accuracy of the SN location determination. The ALMA astrometry is phase-referenced to quasars with well-determined locations, and we do not expect the absolute astrometric calibration of the ALMA data to represent a limiting factor at the $\sim 1''$ scales we work at (see ALMA Technical Handbook). Instead the $1{-}1.5''$ resolution of the ALMA data will usually represent the limiting factor, though for the older half of our sample the location of the SNe may also contribute uncertainty. The OSC does not include positional uncertainty estimates, so to gauge the level of uncertainty we examined the original reference papers (listed in Table \ref{tab:OSCSample}) to estimate the typical positional uncertainty. Many papers do not report a positional uncertainty \citep[e.g.,][]{Martin2005, Marples2016, Kendurkar2018, Tonry2020} though reporting uncertainties has become more common for SN that have occurred in the last $\sim20$ years. When positional uncertainties are reported, they tend to range from 0.1 to 1'' \citep[e.g., see][]{Evans2003, Monard2008, Pignata2009}. Based on this, we assume that for modern SN searches like ASAS-SN \citep{Shappee2014, Kochanek2017}, $\lesssim 1''$ represents a typical uncertainty on the position of a relatively recent SN without extensive follow up observations, meaning that we expect the SNe to be reasonably well determined relative to the ALMA resolution.  However for early SNe, like SN1923A or 1935C in our sample, positional uncertainties could be as high as $\sim10''$. We consider that for the $\sim40\%$ (27/63) of our sample that has occurred within the last $\sim20$ years, the positional uncertainty is better than the angular resolution of PHANGS–ALMA CO maps, while for the remaining $\sim60\%$ (36/63), the uncertainty in the astrometry is dominated by the uncertainty in the reported SN location. We proceed, noting that this situation could be improved by more ALMA observations focused on modern SNe with well-determined locations or by simply waiting for more SNe to explode in the PHANGS targets.

For each SN in the PHANGS--ALMA footprint, we measure the line-integrated CO~(2-1) intensity, $I_{\rm CO}$ expressed in the units of K~km~s$^{-1}$, at the pixel location of the SN in the PHANGS--ALMA ``broad'' moment 0 map and its associated statistical uncertainty. The ``broad" mask is the PHANGS high completeness data product, expected to include all emission but with somewhat higher noise. \citealt{Leroy2021} describe the masking scheme and the procedure used to generate uncertainty maps. Briefly, the mask includes any regions of the data cube detected at either coarse or high resolution. The associated uncertainty map is calculated based on a three-dimensional noise model derived from the empty regions of the data cube and then converted into the map following standard error propagation. The PHANGS--ALMA survey\footnote{PHANGS--ALMA survey data is available at https://almascience.eso.org/alma-data/lp/PHANGS.} produced broad moment 0 maps at fixed physical resolutions of 60, 90, 120, 150, 500, 750, and 1000~pc, as allowed by the angular resolution of the data and the distance to the targets. We report our CO~(2-1) intensity measurements at the pixel location of the SN sites and their statistical uncertainty in Table \ref{tab:intensities} at resolutions of 60, 90, 120, and 150~pc.

This $I_{\rm CO}$ can be translated to a mass surface density of molecular gas, $\Sigma_{\rm mol}$ expressed in the units of  $M_\odot \, {\rm pc}^{-2}$, via:

\begin{equation}
\label{eq_sigmol}
    \Sigma_{\rm mol} = I_{\rm CO}^{2-1} \cdot \alpha_{\rm CO}^{2-1} \cdot \cos{i},
\end{equation} 
where $I_{\rm CO}$ is the integrated CO~(2-1) line intensity expressed in the units of K~km~s$^{-1}$, $\alpha_{\rm CO}$ is the CO~(2-1) to H$_2$ conversion factor, which represents the ratio of the gas mass-to-CO~(2-1) luminosity and has units of [$M_{\odot}$ (K km $\rm s^{-1}$ $\rm pc^{-2})^{-1}]$, and \textit{i} is the inclination of the host galaxy \citep[see][for a discussion of the need to apply an inclination correction even at these scales]{Sun2022}. 

We adopt the metallicity-dependent $\alpha_{\rm CO}^{1-0}$ values calculated in \cite{Sun2020} which uses local metallicities predicted from scaling relations \citep{Sanchez2014, Sanchez2019} and converts them to $\alpha_{\rm CO}^{1-0}$ via a simple power-law scaling (similar to \citealt{Accurso2017}). We convert from $\alpha_{\rm CO}^{1-0}$ to $\alpha_{\rm CO}^{2-1}$ with a CO~(2-1)-to-CO~(1-0) line ratio value, $R_{21} = 0.65$ \citep{DenBrok2021, Leroy2021b}. Because the lower resolution of the data led it to be omitted from the \citet{Sun2020} calculations, we lack an $\alpha_{\rm CO}^{2-1}$ estimate for NGC1068; instead we use the standard value of 6.7 $[M_\odot$ (K km $\rm s^{-1}$ $\rm pc^{-2})^{-1}]$ prescribed in \cite{Bolatto2013}. We calculate the total molecular mass within each resolution element, $M_{\rm mol}$, via:
\begin{equation}
\label{eq_molMass}
    M_{\rm mol} = \Sigma_{\rm mol} \cdot \left(\frac{\theta}{2}\right)^2 \cdot \frac{\pi}{{\rm ln}(2)},
\end{equation}
where $\Sigma_{\rm mol}$ is the molecular gas surface density expressed in units of $M_\odot \, {\rm pc}^{-2}$, and $\theta$ is the FWHM beam size of the CO~(2-1) map in units of parsecs.

We calculate a representative $3\sigma$ $\Sigma_{\rm mol}$ and molecular gas mass measurement for our sample to highlight the general sensitivity of the CO~(2-1) maps. We do this by measuring each $3\sigma$ $\Sigma_{\rm mol}$ value from all of the CO~(2-1) maps in our sample -- accounting for each galaxy's inclination -- and then taking the median value. We do the same for our $3\sigma$ molecular gas mass value. Note that because we account for the galaxies' inclinations when calculating $\Sigma_{\rm mol}$, there is not a single one-to-one mapping between molecular gas mass and $\Sigma_{\rm mol}$ that applies across the entire sample. We measure a median $3\sigma$ $\Sigma_{\rm mol}$ of $7.9~M_\odot {\rm pc}^{-2}$ and a median $3\sigma$ molecular gas mass of $10^{5.3} M_\odot$ at 150~pc resolution.

In \S \ref{sec:MCdistance} we generate model SN populations from the probability distributions of the gas and stellar populations of our galaxies. The density of stellar populations is pulled from 3.6~$\mu$m maps from the Infrared Array Camera (IRAC) on \textit{Spitzer}. The provenance of the IRAC maps is described in \citet{Querejeta2021}; most come from S$^{4}$G \citep{Sheth2010}. We used 3.4~$\mu$m Wide-field Infrared Survey Explorer (WISE) maps \citep{Leroy2019} when IRAC maps are unavailable (for galaxies: Circinus, NGC2997, NGC4945, NGC5128, NGC5530, and NGC6744).



\section{Results}
\label{sec:Results}

Our final cleaned sample consists of 63 SNe within 31 galaxies, each marked with a checkmark in Table \ref{tab:OSCSample}. Overall, our SNe host galaxies have higher mass and higher SFR than the typical PHANGS--ALMA target. The median galaxy with a SNe has $\log_{10} M_\star [{\rm M_\odot}] \sim 10.52$ and $\log_{10} {\rm SFR} [ {\rm M_\odot~yr^{-1}}] \sim 0.36$ while the median PHANGS--ALMA galaxy has $\log_{10} M_\star [{\rm M_\odot}] \sim 10.35$ and $\log_{10} {\rm SFR}  [{\rm M_\odot~yr^{-1}}] \sim -0.09$. This appears consistent with the idea that the SNe rate will broadly trace the SFR or stellar mass of a galaxy, making the more massive, more actively star-forming targets more likely to host a SNe, but the incompleteness of the OSC prevents any formal rate calculation.

Table~\ref{tab:types} shows the distribution of the different SNe types in our sample. We divide our SNe into the following SN type classifications: SNe Ia, SNe II, Stripped-Envelope Supernovae (SESNe), and unclassified SNe. The SESNe consist of SNe Types Ib, Ic, and IIb. The unclassified SNe are either reported in the OSC as ``unclassified" or as Type I SNe. The latter classification of Type I means that there were no hydrogen features in the spectrum used to classify the SN, but this does not provide enough information to distinguish between a CCSN or SN Type Ia.


\begin{deluxetable}{lccccc}[t!]    
\tabletypesize{\small}
\centering
\tablecaption{Our Supernova Sample}
\label{tab:types}
\tablehead{
\colhead{Type} & \colhead{Number} & \colhead{\% of Sample} & \colhead{\% of Classified}} 
\startdata
SNe II & 34 & 54 & 61 \\
SNe Ia & 13 & 21 & 23 \\
SESNe (IIb/Ib/Ic) & 9 & 14 & 16 \\
Unclassified & 7 & 11 & -- \\
All types & 63 & 100 & -- \\
\enddata
\end{deluxetable}


\begin{figure*}[]
    \centering 
    \subfloat[]{%
        \includegraphics[width=0.95\linewidth]{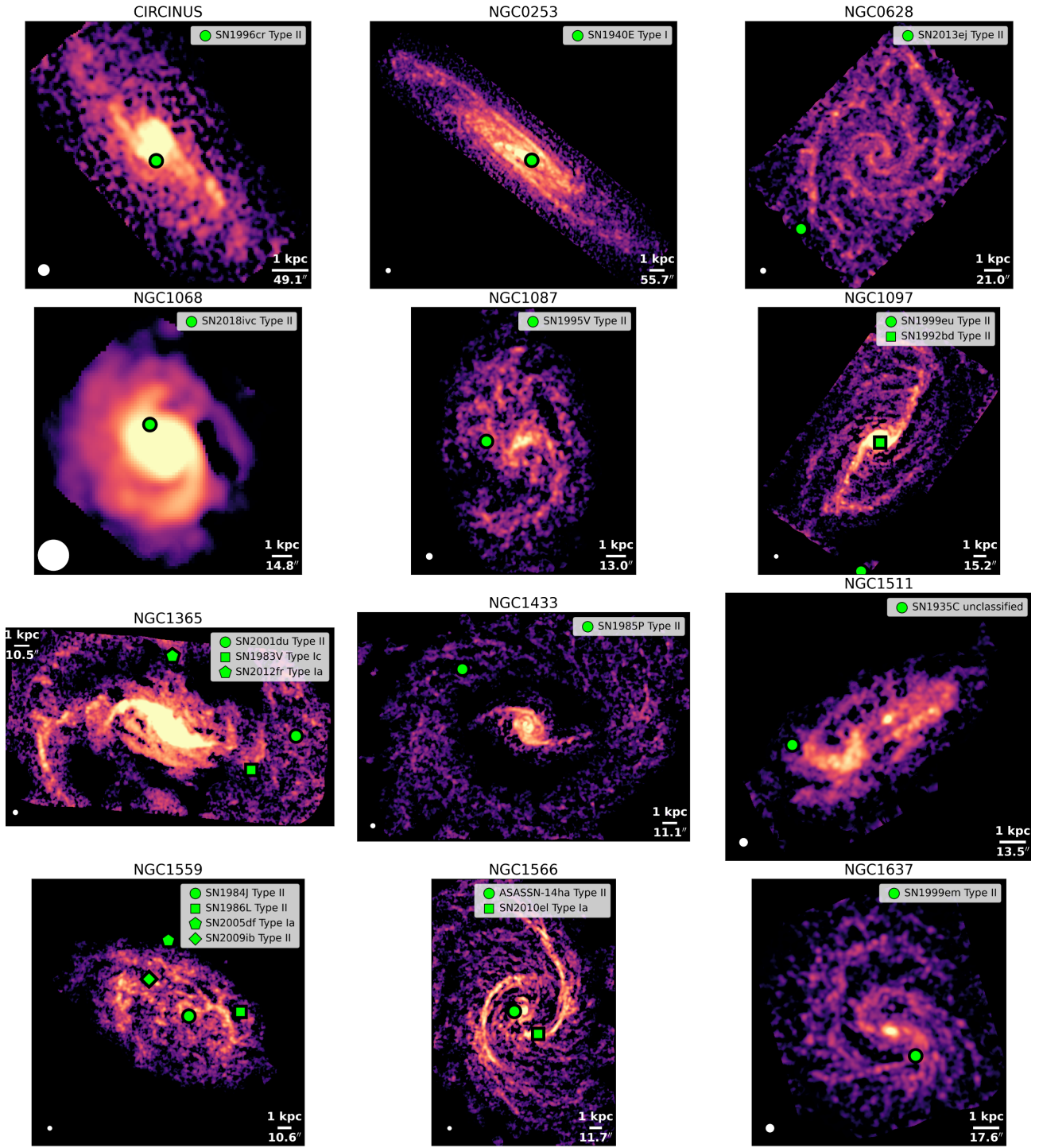}
        \label{fig:1.a}%
    }
\caption{\textit{PHANGS--ALMA CO~(2-1) emission intensity maps for the galaxies with SNe in our sample.} All but three galaxies are shown at 150~pc resolution (NGC1068 is at 750~pc, NGC1672 and NGC4579 are at 500~pc). White contours enclose intensity measurements with signal-to-noise ratios of three or greater. SNe are marked with lime symbols, and different shapes signify each SN within a single galaxy. In the legend, SNe are labeled with their designation and type classification. The beam size of the telescope configuration is marked with a white circle on the bottom left of each plot. Each plot includes a 1 kpc scale bar, with the corresponding angular size in arcseconds. Each panel is oriented with the top of the figure as North, East is left}.
\end{figure*}
\begin{figure*}[ht]\ContinuedFloat                               
\subfloat[]{%
        \includegraphics[width=0.95\linewidth]{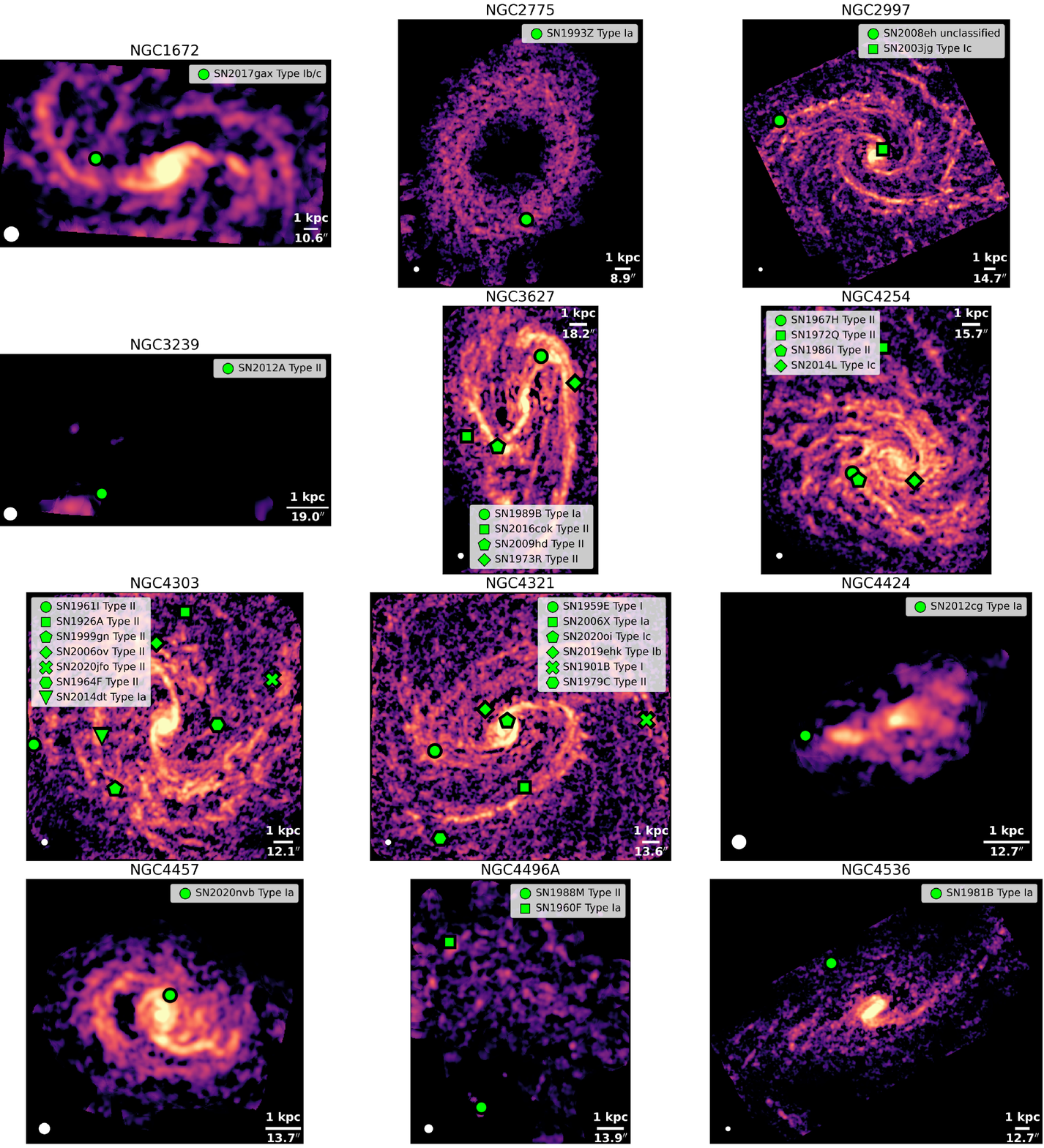}
        \label{fig:1.b}%
        }
\caption{Continued}
\end{figure*}
\begin{figure*}[ht]\ContinuedFloat    
    \subfloat[]{%
        \includegraphics[width=0.95\linewidth]{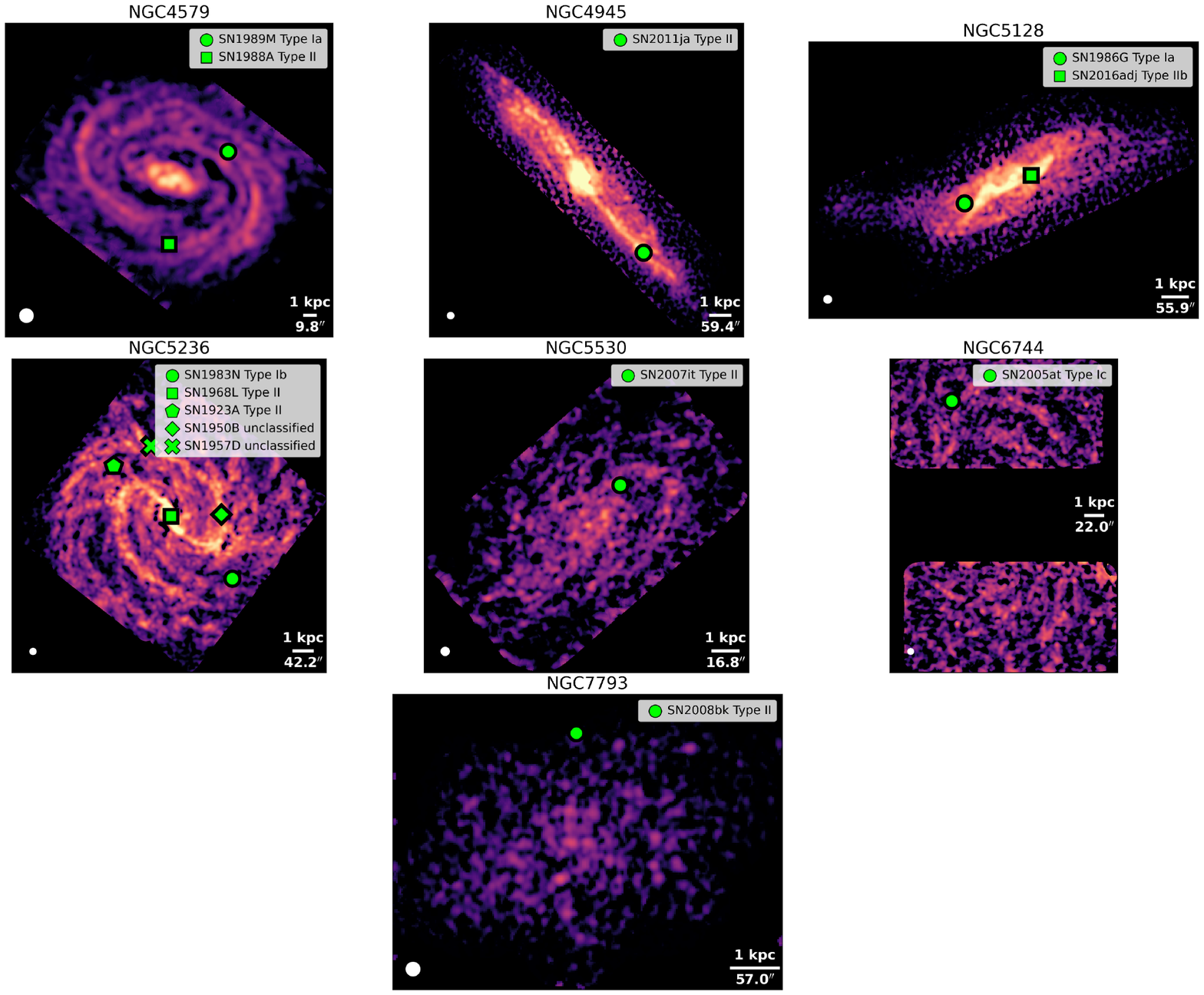}
        \label{fig:1.c}%
}
\label{fig:AllGalaxies}
\caption{Continued}
\end{figure*}

Figure \ref{fig:AllGalaxies} shows the PHANGS--ALMA CO~(2-1) emission maps of the galaxies that host the SNe in our sample. The legends report the designation of the SN and its type classification. 

\subsection{Detection of CO~(2-1) at the location of recent supernovae}
\label{sec:detectionallres}

\begin{figure*}[]
    \centering
    \includegraphics[width=1.0\textwidth]{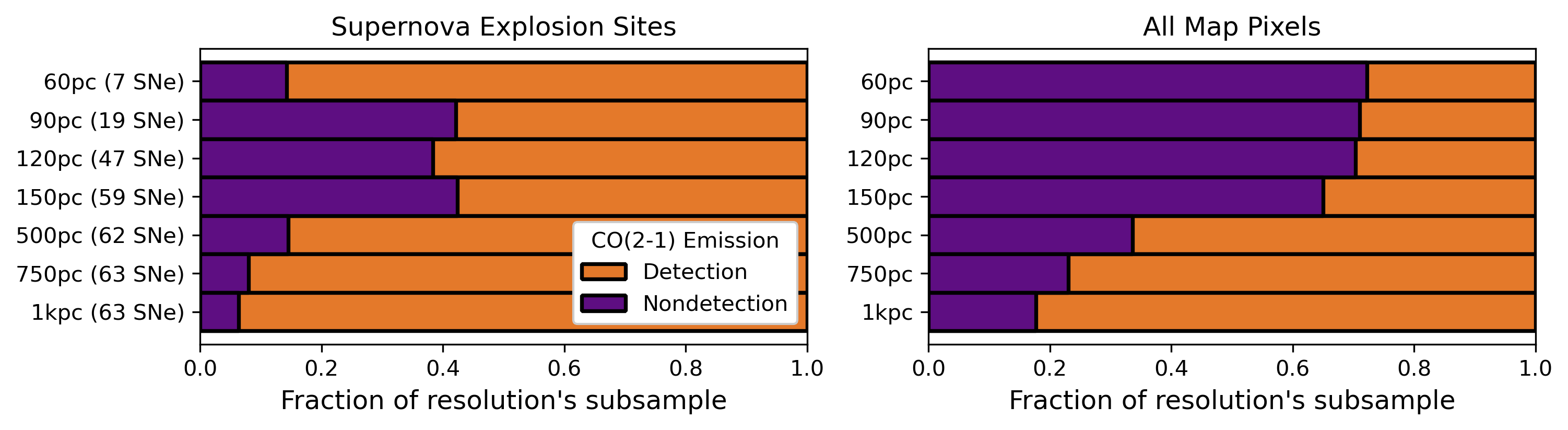}
    \caption{\textit{Fraction of CO~(2-1) non-detections (purple), and detections (orange) at resolutions from 60 to 1000~pc for our SNe explosion sites (left panel) and for all pixels in the CO~(2-1) maps for our sample (right panel)}. The number of SNe in the sample at each resolution is listed in parentheses on the $y$-axis of the left panel. The number of SN decreases as the resolution increases because we have fewer maps available at higher resolution. We classify a signal-to-noise $\geq$ 3 to be a detection. At 1~kpc resolution almost all (59/63) SNe in PHANGS--ALMA show associated CO~(2-1) emission. By 150~pc this drops to roughly even fractions of detections and non-detections. Comparing these sites to all map pixels, we find that our SN explosion sites tend to occur in regions of higher CO~(2-1) emission than if they were randomly drawn from any pixel in the map.}
    \label{fig:SNRAcrossResolutions}
\end{figure*}

First, we measure how frequently CO~(2-1) emission is detected coincident with the sites of recent SNe. To do this, we use the signal-to-noise ratio measured from the PHANGS--ALMA intensity maps described in \S \ref{sec:Methods}. For this exercise, we define signal-to-noise $\geq3$ as a detection of molecular gas. We repeat this exercise for multiple resolutions, from 60~pc to 1~kpc. We show the results of these measurements in Figure~\ref{fig:SNRAcrossResolutions} and Table~\ref{tab:SampleAcrossResolutions}. 

We detect CO~(2-1) emission at $\sim95\%$ (59/63) of the SN locations at 1~kpc resolution. Out of the 63 SNe, four\footnote{SN1988M in NGC4496A, SN1999eu in NGC1097, SN2005df in NGC1559, SN2012A in NGC3239} do not have CO~(2-1) detections coincident with their locations on kpc scales. These four are near map edges, and thus their non-detections are ambiguous. Finding CO~(2-1) emission at the location of almost every SN at 1 kpc resolution may not be surprising. The PHANGS--ALMA survey targets star-forming galaxies, the survey area was chosen to match the extent of infrared (IR) emission in the galaxy, and the PHANGS--ALMA maps have very good surface brightness sensitivity at $\sim1$~kpc resolution \citep[see][]{Leroy2021}. Accordingly, in \citet{Sun2022}, $\sim80\%$ of the 1.5 kpc-size regions covering the complete PHANGS--ALMA survey show some detectable CO~(2-1) emission. Thus, at $\sim1$~kpc resolution, our observations mostly show that we are studying the the gas-rich, star-forming parts of star forming galaxies.

As the resolution improves, the fraction of CO~(2-1) detections at the locations of recent SNe decreases. This reflects the combination of two effects: 

\begin{enumerate}
\item[1.] At coarser resolutions, a beam will be more likely to cover both a SN and nearby gas clouds that may not be directly coincident or physically associated with the SN. Scales of 150-500~pc correspond roughly to the range of distances between individual molecular gas peaks in PHANGS--ALMA \citep[][J. Machado et al. in prep]{Chevance2020,Kim2022}, so that at these resolutions it becomes increasingly likely to find a CO~(2-1) peak within a randomly placed beam.

\item[2.] The RMS noise of the CO~(2-1) maps increases as the resolution becomes sharper. This means that while faint, diffuse CO~(2-1) emission could be detected on 1~kpc scales, only gas with higher surface or column densities will be detected on $\sim$50--150 pc scales. 
\end{enumerate}

In massive ($M_{\star} \gtrsim 10^{10} M_{\odot}$), CO-bright galaxies, the first effect is likely dominant, because the PHANGS--ALMA CO~(2-1) maps have high completeness\footnote{Completeness is the ratio of a map's flux that meets the signal-to-noise cutoff criteria to the map's total flux. For the subset of PHANGS--ALMA CO~(2-1) galaxies in our sample at 150~pc resolution, we have an overall average completeness value of 70\% \citep[see][]{Leroy2021} with an average completeness of $75\%$ for the the higher-mass half of the sample and 60\% for the lower-mass half of the sample.}, even at high resolution. In lower-mass ($M_{\star} \lesssim 10^{10} M_{\odot}$) galaxies where CO~(2-1) becomes fainter, the increased noise at high resolution may make a more significant difference, as the completeness of the PHANGS-ALMA maps at high resolution is worse for low-mass galaxies. 

The picture is more complex at even higher resolution. 150~pc is the best common linear resolution achievable for all but three (NGC1068, NGC1672, and NGC4579) of the PHANGS--ALMA galaxies in our sample. Therefore, we consider 150~pc to be our most comprehensive resolution, covering $\sim95\%$ (59/63) of our SNe sample. At this resolution, one beam corresponds to roughly the scale of an individual GMC or a giant molecular association, and we detect CO~(2-1) emission from $\sim60$\% of our SN sample. At 150-pc resolution, we do not detect CO~(2-1) emission at $\sim40$\% of the SN locations. This is a significant increase in non-detections relative to the 1-kpc case. These non-detections are assigned an upper-limit CO~(2-1) intensity measurement of 3 times the noise value at their location. We report our CO~(2-1) intensity measurements and their statistical uncertainty in Table \ref{tab:intensities}. In \S \ref{sec:detection150pc} we use the 150~pc maps to compute molecular gas masses and other properties at the locations of our SNe.

Figure \ref{fig:SNRAcrossResolutions} implies that $\sim95\%$ of SNe in the PHANGS--ALMA footprint explode within 1~kpc of detectable molecular gas and $\sim60\%$ explode coincident with molecular gas emission at 150~pc resolution or finer. The apparent increase in detections at the SN sites at 60~pc resolution is likely due to the reduction in sample size than due to an underlying physical cause.


\begin{deluxetable}{lccccc}[tbp]
\label{tab:SampleAcrossResolutions}
\centering
\tabletypesize{\small}
\tablecaption{CO~(2-1) Detection Rates at SN Sites}
\tablehead{
\colhead{Resolution} & \colhead{Number of SNe} &  \colhead{\% Detections}}
\startdata 
60~pc & 7 & 86 \\
CCSNe only & 5 & 80 \\
\hline
90~pc & 19 & 58 \\
CCSNe only & 13 & 54 \\
\hline
120~pc & 47  & 62 \\
CCSNe only & 31 & 65 \\
\hline
150~pc & 59 & 58 \\
CCSNe only & 40 & 60 \\
\hline
500~pc & 62 & 85 \\
CCSNe only & 42 & 88 \\
\hline
750~pc & 63 & 92 \\
CCSNe only & 43 & 91 \\
\hline
1000~pc & 63 & 94 \\
CCSNe only & 43 & 93 \\
\enddata
\end{deluxetable}

This has two implications. First, the fact that $\sim60\%$ of SNe \textit{do} appear coincident with molecular gas at 150~pc resolution suggests that interactions between SNe and MCs could be common. These interactions have the potential to destroy GMCs and shut down star formation in these areas \citep{Walch2015,Keller2020,Lu2020} or they may trigger future generations of MC or star formation when the shock wave compresses nearby clouds \citep[e.g.,][]{Zucker2022}.

Second, the fact that $\sim40\%$ of SNe \textit{do not} appear coincident with molecular gas suggests these explosions have the potential to exert significant feedback on the diffuse ISM, stirring turbulence on large scales, heating the diffuse gas, and even ejecting material out of the galaxy \citep{Martizzi2016, Andersson2020, Lopez2020}. 

The first result qualitatively agrees with observations of SNRs in the Milky Way, though the quantitative implications are less clear. Interactions between SNRs and MCs are seen relatively frequently in the Milky Way. For example, $\sim22\%$ of the 294 known Galactic SNRs \citep{Green2019} have signatures of interaction with surrounding molecular gas \citep{Jiang2010}. Quantitatively, the exact frequency of such interactions are less clear due to the heterogeneous compilation of SNRs and non-uniform selection effects in \cite{Green2019}, and Milky Way studies achieve better mass sensitivity than our extragalactic CO~(2-1) maps (i.e., they are not restricted to detecting only GMCs). Attempting a more systematic survey of SNR--MC interactions, \cite{Kilpatrick2016} placed a 37\% upper-limit on SNR--MC interactions but studied only a subset of the remnant population within the Milky Way. Given that the remnants produced by individual SN explosions typically reach a maximum diameter of $\sim 50$ pc (see \S \ref{sec:zooms}), our observation of $\sim60\%$ of SNe coincident with molecular gas at 150~pc resolution appears consistent with the local result. This also highlights the caveat that an explosion that appears coincident at this resolution still may not interact with nearby molecular gas during its remnant phase. We return to this in \S \ref{sec:zooms}.

The second result agrees well with the recent observations of \textsc{Hii} region-GMC separations and \textsc{Hii} region evolution discussed in \S \ref{sec:intro}. These observations imply a strong role for pre-SN feedback in clearing the immediate molecular gas near sites of star formation. Our observations show directly that for $\sim40\%$ of all SNe there is no detectable massive CO-emitting cloud near the explosion site. Our observations do not rule out the presence of lower-mass MCs.


\subsection{Detection properties at 150~pc resolution}
\label{sec:detection150pc}

In the previous section, we looked at the presence or absence of molecular gas. Now we assess the molecular gas mass associated with each of our SNe and examine how the intensity of CO~(2-1) emission at the sites of SN explosions compares to the distribution in the galaxy as a whole.

\begin{figure*}[hbtp]
    \centering
    \includegraphics[width=1.0\textwidth]{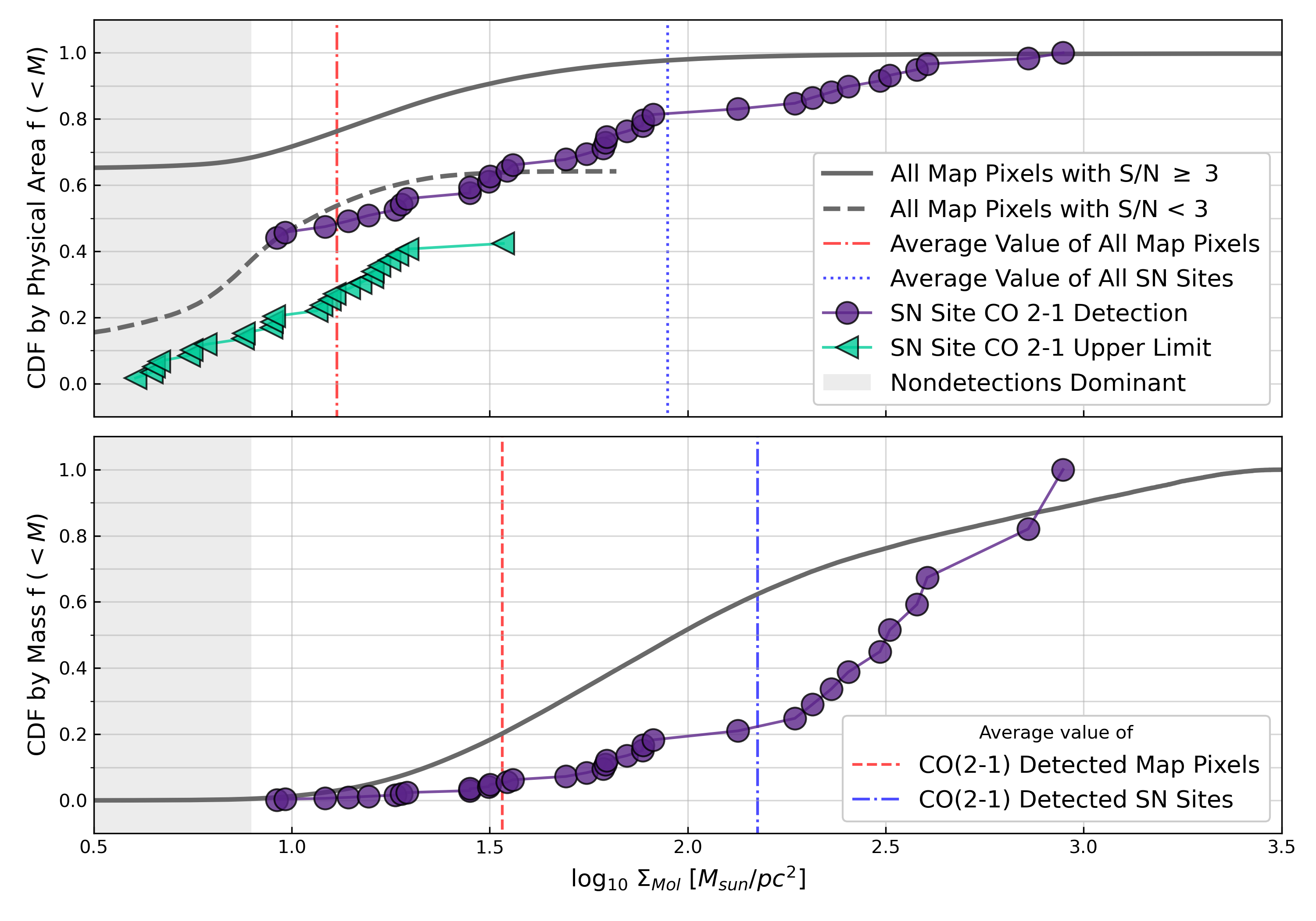}
    \caption{\textit{CDFs of physical area (top panel) and molecular gas mass (bottom panel) as a function of projected molecular gas surface density at the locations of our SN sample at 150~pc resolution.} We classify the measurements into detections with signal-to-noise $\geq$ 3 (purple circles) and non-detections plotted at the upper-limit of three times the noise value of the measurement (turquoise triangles). The top panel weights each pixel by its physical area in $pc^2$ in the plane of the galaxy, while the bottom panel weights each pixel by the molecular mass within the pixel. In the top panel we generate the CDF using both non-detections and detections. When calculating the CDF, we assume that the non-detections all have lower $\Sigma_{\rm mol}$ than the lowest detection value. The bottom panel omits non-detections, assuming the pixels to have negligible molecular gas. For comparison, we also plot the CDF of all of the pixels in all the CO~(2-1) maps in our sample, classifying them by signal-to-noise with dark-grey, solid (detections) and dark-grey, dashed (upper-limits) lines. Finally, we plot the average value of the map pixels with a red, vertical line and the average value of the SN sites with a blue, vertical line. Both panels show that SNe consistently explode in regions of brighter CO~(2-1) emission and molecular gas surface density compared to a pixel drawn randomly from the map.}
    \label{fig:CDF_ALLSNe}
\end{figure*}

\begin{figure*}[]
    \centering
    \includegraphics[width=1.0\textwidth]{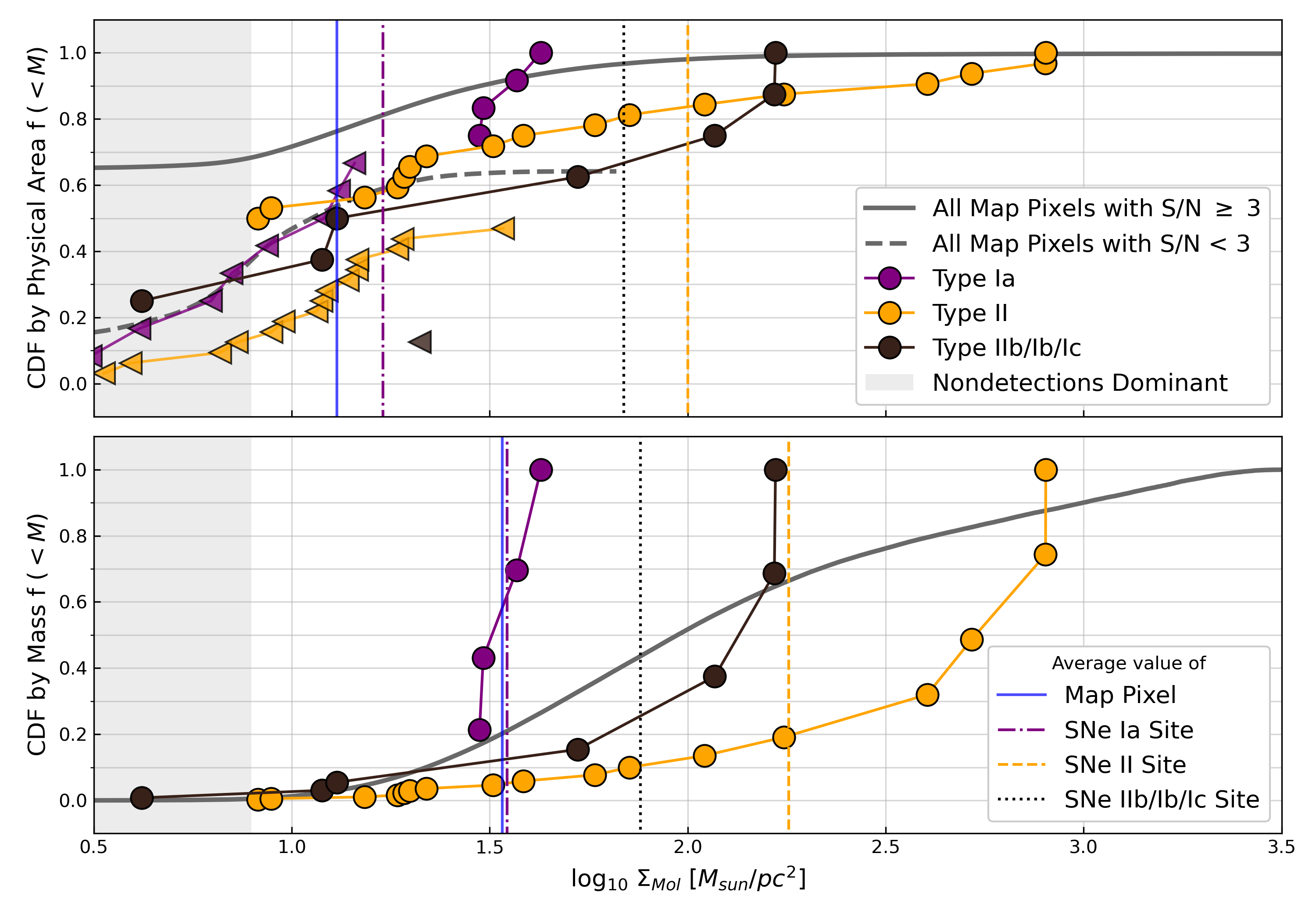}
    \caption{\textit{CDFs of physical area (top) and molecular gas mass (bottom) at the locations of SNe as a function of molecular gas surface density, now sorted by type.} As Figure \ref{fig:CDF_ALLSNe} but now separating the SNe by type. Upper-limit values are marked with triangles, and detections are marked with circles. SNe Ia have purple markers, SNe II have orange markers and SESNe (SNe IIb, Ib, and Ic) have black markers. In the top panel, the CDFs of all the pixels in the galaxy maps are also plotted, with detections marked by a solid, dark-grey curve and non-detections (with upper-limit values) with a dashed, dark-grey curve. We plot the average surface density value of map pixels with a solid, blue, vertical line; and the average surface densities at the sites of SNe Ia, SNe II, \& SESNe (IIb/Ib/Ic) with purple-dot-dashed, black-dotted, and orange-dashed vertical lines respectively. In the bottom panel, we plot the fraction of \textit{molecular gas mass} instead of the fraction of physical area. The average surface density of map pixels with detected CO~(2-1) emission is plotted with a solid, blue, vertical line and the average surface density of Type Ia, II, and IIb/Ib/Ic sites with detected CO~(2-1) emission are plotted with purple-dash-dotted, orange-dashed, and black-dotted vertical lines respectively. The figure shows that SESNe and Type II SNe tend to occur in regions of higher $\Sigma_{\rm mol}$  than Type Ia SNe. We also note that $\sim35\%$ of our SNe Ia occur in regions with detectable molecular gas, while $\sim65\%$ of our SNe Ia occur with no detectable CO~(2-1) emission.}
    \label{fig:CDFByType}
\end{figure*}

To do this, we convert the CO~(2-1) intensity to molecular gas surface density$, \Sigma_{\rm mol}$, as described in \S \ref{sec:Methods}. Then we compute distributions of $\Sigma_{\rm mol}$ at the locations of our SN sample and for each pixel in the host galaxies' 150~pc CO~(2-1) maps. Figures \ref{fig:CDF_ALLSNe} and \ref{fig:CDFByType} show the results. 


\subsubsection{Detection properties for all SNe}

In Figure \ref{fig:CDF_ALLSNe} we plot the cumulative distribution function (CDF) of physical area (top panel) and molecular gas mass (bottom panel) as a function of molecular gas surface density estimated from CO~(2-1) emission. The plot combines all 59 SNe from 29 host galaxies that have PHANGS--ALMA 150~pc CO~(2-1) maps.

In the top panel we distinguish between detections for 36 of our SNe and non-detections for the remaining 23. For non-detections, we plot that pixel's $3\sigma$ upper-limit value as explained in \S \ref{sec:detectionallres}. We weigh each pixel by the physical area in the plane of the galaxy. In the bottom panel, we show only results for detections, assuming non-detections to have zero associated molecular gas mass. Here we weigh each pixel by the amount of estimated gas mass within the pixel. In both panels, the dark-grey lines show the CDF derived from all pixels in all of the CO~(2-1) maps for galaxies with a SN in our sample.

Figure \ref{fig:CDF_ALLSNe} shows that SNe consistently explode in regions of higher CO~(2-1) intensity and therefore higher molecular mass compared to a typical pixel in a PHANGS--ALMA CO~(2-1) map. Compared to the full PHANGS--ALMA field of view, locations of SN show fewer non-detections and a larger fraction of pixels at higher intensity. For example, the top panel of Figure \ref{fig:CDF_ALLSNe} shows that $\sim60\%$ of our SNe occur where CO~(2-1) has been detected, while $\sim40\%$ of the pixels in the PHANGS--ALMA maps have a detection with signal-to-noise $\geq$ 3. Even considering only detections, the pixels with SNe show higher mass and intensity values at a given percent. That is, the CDF of the SNe locations is consistently shifted to the right relative to the CDF of all pixels.

The top panel of Figure \ref{fig:CDF_ALLSNe} also allows us to phrase our detections in physical terms rather than signal-to-noise. We find that $\sim50\%$ of the SN locations at 150~pc resolution have $\Sigma_{\rm mol} > 10~M_\odot$~pc$^{-2}$ at 150~pc resolution. Although the mass sensitivity varies slightly across the survey due to the varying inclination of the targets, this $\Sigma_{\rm mol}$ corresponds in general to $M_{\rm mol}$ in the beam $> 10^{5.5}$~M$_\odot$. This mass corresponds to the mass of an individual GMC \citep[e.g.,][]{Rosolowsky2021}. Thus, about half of the SNe in our sample explode within about 75~pc (the radius of the 150~pc beam) a the total molecular gas mass equivalent to a medium-mass GMC. For this measurement we do not distinguish between whether this mass is part of a larger structure or the combination of multiple lower-mass clouds.

Our CO~(2-1) detection limit at 150~pc resolution corresponds to a molecular gas surface density of $\sim 7.9 M_\odot {\rm pc}^{-2}$ or $\sim 10^{5.3} M_\odot$ (see \S \ref{sec:Methods}). As the plot shows, the non-detections could still harbor low-mass clouds, on the order of a few times $10^4 M_\odot$. An individual, low-mass cloud is not likely to produce many high-mass stars given the typical SFR-to-gas ratio in galaxies. Instead, we expect the massive stars that produce SNe to be mostly associated with GMCs because such clouds hold most of the molecular gas mass and host most of the star formation in galaxies \citep[e.g., see][]{Rosolowsky2005,Fukui2010, Murray2010}. However, we cannot rule out that ensembles of low-mass clouds might collectively produce SNe. Regardless, we can be confident that most of the molecular gas mass is accounted for due to the completeness of the CO~(2-1) maps, where $\sim70\%$ of the total CO~(2-1) flux in our maps is detected in the $150$~pc resolution maps at signal-to-noise $\geq3$.

In the bottom panel of Figure \ref{fig:CDF_ALLSNe} we look at CDFs calculated from the total molecular gas mass budget. This allows us to test whether the distribution of mass at SNe locations resembles the distribution of molecular gas mass in the maps as a whole. For this calculation, we consider only detected pixels with a signal-to-noise of 3 or greater for both the maps and SNe. We expect this clipping to make little difference because usually only a small amount of the total molecular mass exists in non-detected regions. We weight each pixel by the amount of measured mass. This translates to the fraction of the total molecular gas mass that has a given surface density. The distribution for SNe in Figure~\ref{fig:CDF_ALLSNe} appears displaced significantly to the right of the distribution for all pixels, and has a somewhat shallower slope with no clear saturation or turnoff. This implies that the mass function of CO~(2-1) emitting regions producing visible SNe is overall higher than the general mass function of CO~(2-1) emitting clouds. It also suggests that the brightest regions are particularly likely to be associated with SNe.

SN surveys are less likely to find SNe in regions with high gas surface density, since these are typically located towards galaxy centers and are correlated with higher extinction. Since high gas surface density implies high CO~(2-1) mass, this bias will make it somewhat less likely to find SNe associated with high concentrations of molecular gas. Recent IR-transient surveys, e.g. SPIRITS \citep{Kasliwal2017}, have discovered SNe only at IR wavelengths at $A_{\rm V} \sim $2-8, and estimate that optical surveys could miss about 40\% of CCSNe in nearby starburst galaxies \citep{Jencson2019, Fox2021}. We checked within our own sample and find that 18\% of the well-detected CO~(2-1) emission has CO intensity corresponding to $A_v \gtrsim 8$~mag for a standard dust-to-gas ratio and CO-to-H$_2$ conversion factor. If SNe exactly followed the CO distribution, we would expect to find 18\% ($\sim11/63$) of our supernovae in these high-extinction areas. Instead we find 4 SNe, implying that we might miss $\sim 7$ or $\sim 10\%$ SNe in this simplest case. We caution that we do not expect SNe to exactly follow the CO distribution; the star formation efficiency of molecular gas, the CO-to-H$_2$ conversion factor, and the dust-to-gas ratio may all vary. However, this provides a useful order-of-magnitude estimate that demonstrates how we might miss a modest $\sim 10\%$ subset of SNe that occur in high-extinction regions. Despite this, we find a clear correlation between molecular gas and the locations of SNe, implying that if we were able to account for this bias, our correlation would likely be stronger and the actual fraction of SNe near/inside CO~(2-1) emitting clouds is likely higher than observations suggest.


\subsubsection{Detection properties by SN type}
Figure \ref{fig:CDF_ALLSNe} makes no distinction between SN types. In Figure \ref{fig:CDFByType} we repeat our analysis from Figure \ref{fig:CDF_ALLSNe}, but now we separate our SNe sample into into three groups, Type Ia, Type II, and SESNe (composed of SNe IIb, Ib, and Ic). Again, we plot the CDF of $\Sigma_{\rm mol}$ for the entire set of maps for reference. Other than separating the SNe sample by type, Figure~\ref{fig:CDFByType} follows exactly the same procedure as Figure \ref{fig:CDF_ALLSNe}.

With the caution that this sample size will produce only small number statistics, we also note the fractions of the association of specific SN types with nearby molecular gas. We find that CO~(2-1) emission at 150~pc resolution is detected at the location of $\sim55\%$ (17/32) of the Type II SNe, $33\%$ (4/12) of the Type Ia SNe, $\sim90\%$ (7/8) of the SESNe (SNe IIb/Ib/Ic), and at $\sim85\%$ (6/7) of the untyped SNe. 

The finding that $\sim65\%$ of the Type Ia SNe occur away from molecular gas is in good agreement with the putative scenario where SNe Ia come from older stellar populations that are less associated with clouds of star-forming gas.

The result that only $\sim50\%$ of Type~II~SNe are associated with bright molecular emission also appears to agree with the statements about the importance of pre-SN feedback above. However, we also note that the drift of massive stars away from their parent clouds over the $\sim 20{-}40$~Myr delay time of a typical CCSN and the possibility of forming massive stars in lower-mass clouds may also play important roles in this measurement. Regardless, this observation confirms that CCSNe have the potential to inject significant feedback and energy into the diffuse ISM and that this feedback is not only driven by SNe Ia. 

Difficulty in obtaining spectra in the highest-density regions of galaxies, which will also have high extinction, could explain why many of the SNe found in high-density regions are untyped.

We do see a difference in the CDFs of $\Sigma_{\rm mol}$ between SNe types. While less than 40\% of the map coverage has detectable CO~(2-1) emission, we find that the $\sim90\%$ (7/8) of our SESNe sample have well-detected CO~(2-1) emission -- $\sim55\%$ (4/7) of those occurring in regions of $\Sigma_{\rm mol} > 30~M_\odot~{\rm pc}^{-2}$ and higher. This is distinct from our population of Type II SNe which have detections for $\sim55\%$ (17/32) of the sample, with $\sim60\%$ (10/17) occurring in regions of $\Sigma_{\rm mol} > 30~M_\odot~{\rm pc}^{-2}$ and higher. These results are not surprising: we expect the stars with the shortest lifetimes to be more likely to be found in the densest regions, explaining the preference for the SESNe sites to be more closely associated with dense molecular gas.

We find that $33\%$ (4/12) of our SNe Ia occur in regions rich with molecular gas---100\% (4/4) of these in regions of $\Sigma_{\rm mol} > 30~M_\odot~{\rm pc}^{-2}$ and higher---while the remaining $\sim67\%$ of our SNe Ia occur with no detectable CO~(2-1) emission.

\begin{deluxetable}{ccc}[t!]
\label{tab:150MGSDAverages}
\centering
\tabletypesize{\small}
\tablecaption{Average Molecular Gas Mass Surface Density at 150~pc resolution $[{\rm M}_\odot~{\rm pc}^{-2}]$}
\tablehead{
\colhead{Sample} & \colhead{All} & \colhead{Detections Only}}
\startdata
Map Pixels & 13 & 34 \\
All SNe & 89 & 150 \\
SESNe & 69 & 76 \\
SNe II & 100 & 180 \\
SNe Ia & 17 & 35 \\
Unclassified SNe & 120 & 130 \\
\enddata
\end{deluxetable}

We compare the average molecular gas surface density across each of our subsamples in Table \ref{tab:150MGSDAverages}. The average $\Sigma_{\rm mol}$ value of any pixel in a PHANGS--ALMA map at 150~pc resolution is $13M_\odot~{\rm pc}^{-2}$ and an average $\Sigma_{\rm mol} = 34 M_\odot~{\rm pc}^{-2}$ when considering only pixels with detectable CO~(2-1) emission at signal-to-noise $\geq$ 3. Comparing this to the pixels at SN locations, we find this value increases to an average pixel value of  $\Sigma_{\rm mol} = 89 M_\odot~{\rm pc}^{-2}$ across all locations and  $\Sigma_{\rm mol} = 150 M_\odot~{\rm pc}^{-2}$ for CO~(2-1) detections only. Broken down by SN type, we find that our CCSNe (SNe II and SESNe) occur in much brighter CO~(2-1) emitting regions than our SNe Ia, which have average $\Sigma_{\rm mol}$ values only slightly higher than a random map pixel. The high average $\Sigma_{\rm mol}$ values for the unclassified population of SNe is not surprising because the majority of these SN occur in high-extinction regions of their host galaxies.

\begin{figure}[]
    \centering
    \includegraphics[width=0.5\textwidth]{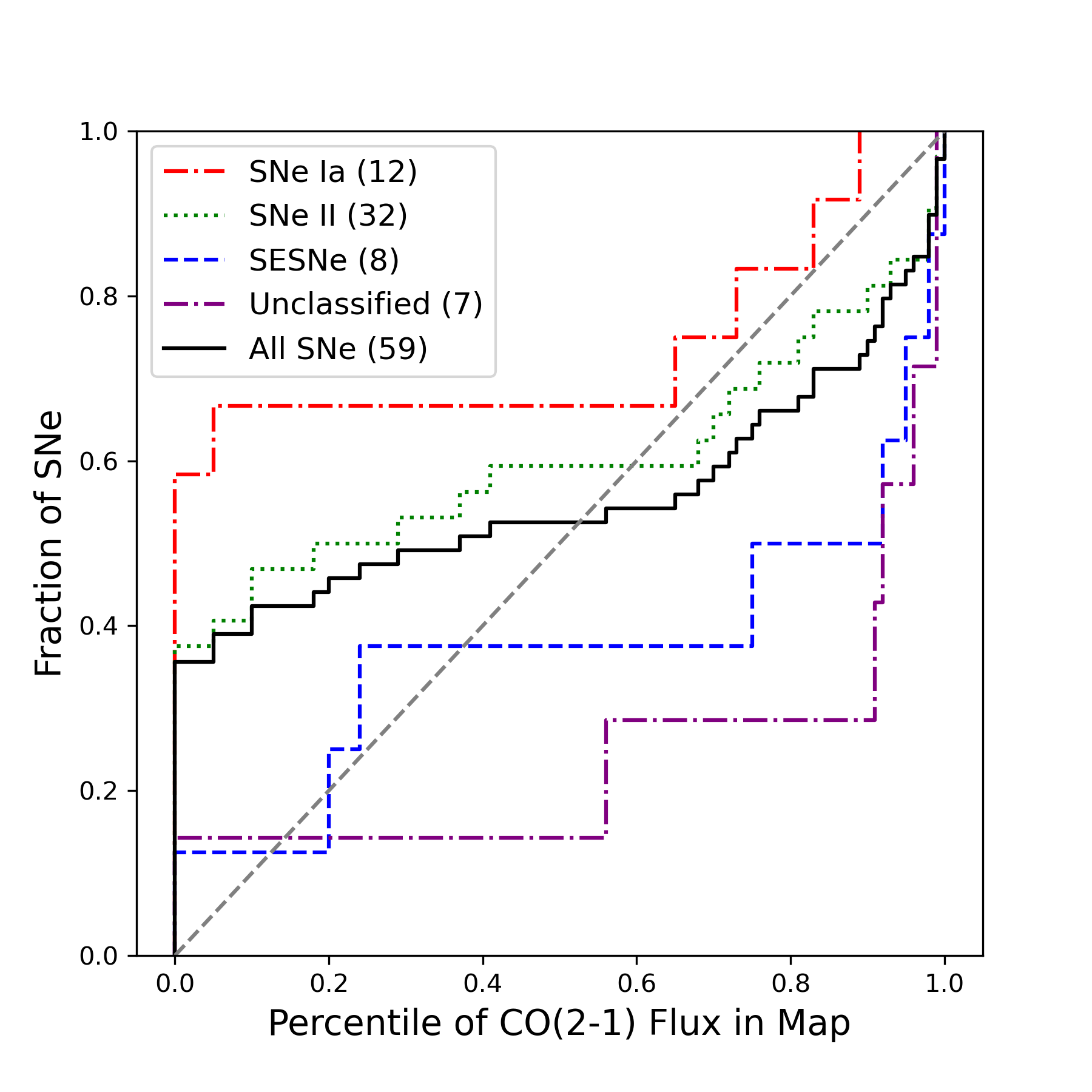}
    \caption{\textit{Normalized cumulative rank (NCR) plot comparing SNe locations to the molecular mass distribution.} Following \citet{James2006} and \citet{Galbany2017} the figure shows the fraction of SNe detected at or below each percentile level in the molecular gas mass map estimated from CO~(2-1). If the SN locations were to perfectly track the molecular gas mass, we would expect the data to follow the one-to-one line. The figure reinforces the results seen above, showing a significant population of SNe Ia, SNe II, and all SNe with little or no CO, but a tendency for SNe with detected CO~(2-1) to be concentrated towards brighter CO~(2-1) emission. SESNe and unclassified show widespread detections and a concentration towards the brightest CO~(2-1) emission.
    }
    \label{fig:CDFAllTypes}
\end{figure}

As a final view of the joint distribution of SNe and CO~(2-1)  emission, Figure \ref{fig:CDFAllTypes} shows the normalized cumulative rank (NCR) of SNe relative to the distribution of molecular gas mass in the CO~(2-1) maps. \citet{James2006} give an excellent overview of the NCR method. Briefly, we sort the pixels in each CO~(2-1) map according to the molecular gas mass in that pixel. Then we note the percentile of the pixel at which each SN occurs. In the case where SNe locations trace the distribution of molecular gas mass, we would then expect a one-to-one line in Figure \ref{fig:CDFAllTypes}. That is, this plot shows how well the locations of SNe follow the distribution of CO~(2-1) intensity within galaxies. A relation above the line indicates that the SNe preferentially occur at low CO~(2-1) intensity; a relation below the line indicates that SNe are concentrated to higher CO~(2-1) intensities. 

Figure \ref{fig:CDFAllTypes} shows that although overall the SN locations are broadly associated with CO~(2-1) emission, none of the SNe types exhibit a perfect 1-to-1 tracing of CO~(2-1) emission. Instead, the NCR plot illustrates many of the points we have seen earlier in this section. SNe Ia and II both show a substantial fraction of SNe associated with little or no CO~(2-1) mass (the high values at low CO~(2-1) percentiles), while SESNe and unclassified SNe mostly show CO~(2-1) detections and appear near bright CO~(2-1) emission. Meanwhile at high intensities, SNe II, SESNe, unclassified SNe, and all SNe all appear concentrated at high CO~(2-1) intensities relative to the overall maps.


\subsection{Comparison of gas properties at 150~pc to those at finer resolutions}
\label{sec:detectionFiner}

In \S \ref{sec:detection150pc} we characterized the gas at the sites of SN explosions at 150~pc resolution, which is the best common resolution for our sample. But at this resolution, there is still uncertainty as to whether the SN will eventually interact with the molecular gas in the beam. Lower energy SNe or those in dense environments may never expand far enough to reach the molecular gas before fading, especially if the molecular gas is not centrally located. Fortunately many of our maps reach higher resolution, and we can look to finer resolutions to better measure where the SNe are in relation to the peaks of the molecular gas surface density on smaller scales.

\begin{figure}[hbtp]
    \centering
    \includegraphics[width=0.5\textwidth]{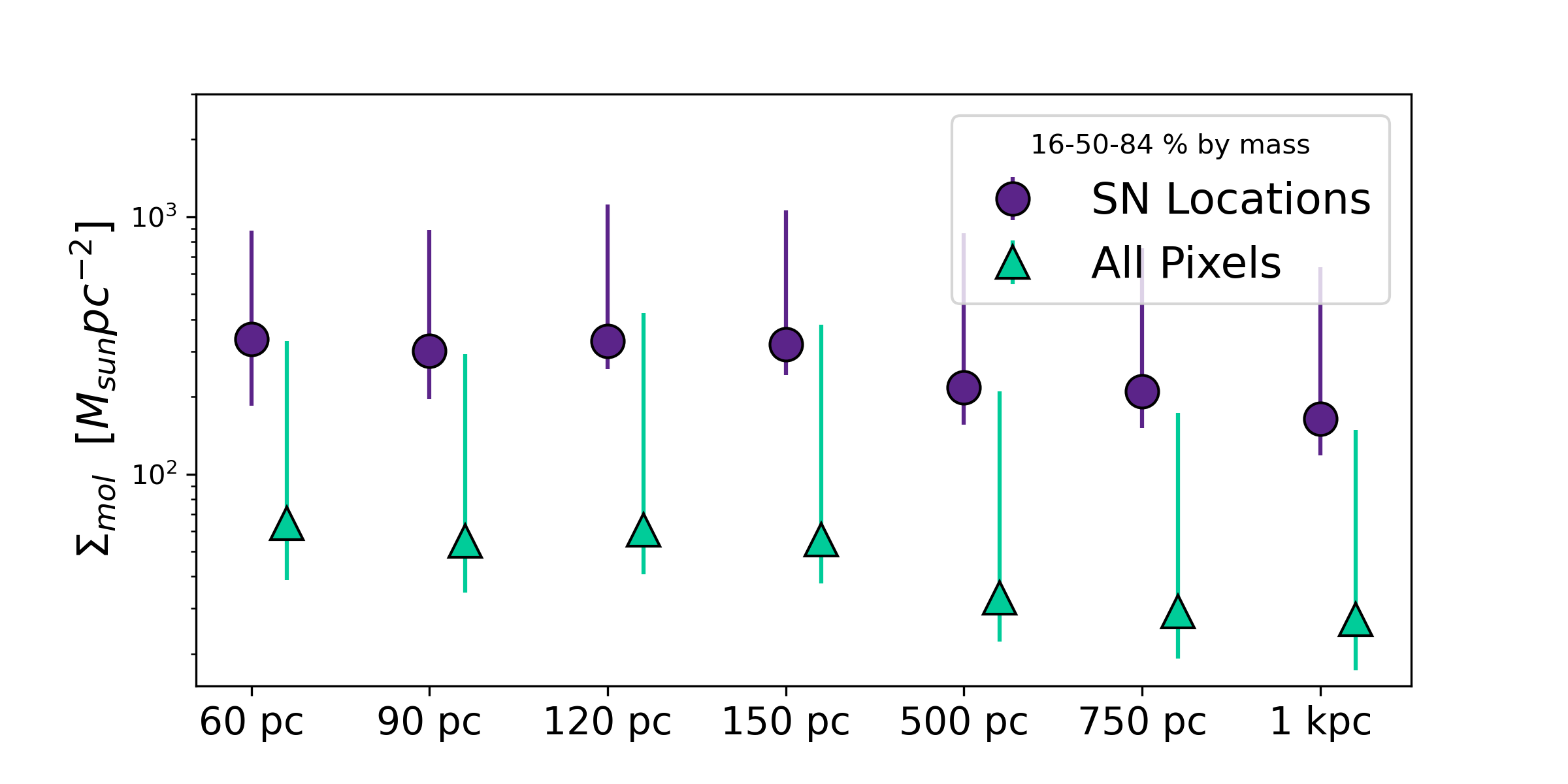}
    \caption{\textit{Mass weighted molecular gas surface density distribution at all resolutions} for SN locations (dark-purple circles) and all pixels (turquoise triangles). Error bars range from the 16th-84th percentile values for each and markers are placed at the 50th percentile measurement. The triangles are offset to a slightly larger x to prevent the error bars from overlapping.}
    \label{fig:MGSDbyRes}
\end{figure}

As a first step, in Figure \ref{fig:MGSDbyRes} we compare the molecular gas surface density distribution, weighted by mass, for both the SN explosion sites (dark-purple circles) and the full set of pixels in the CO~(2-1) maps (turquoise triangles) for all galaxies in our sample. We plot the 16th-84th percentiles for each resolution, with a marker placed at the 50th percentile. We report the specific values for the $16^{\rm th}$, $50^{\rm th}$, and $84^{\rm th}$ percentiles by mass in Table \ref{tab:MGSDresolutions}. Recall (\S \ref{sec:detectionallres}) that the number of maps available decreases as we go to higher resolution, so the amount of data entering each measurement varies.

We find that regardless of resolution, SN tend to occur in regions of molecular gas surface density approximately an order of magnitude greater than the map pixels in general. The median percentile value of the SN explosion sites are on the order of the 84th percentiles from the map pixels.

\begin{deluxetable}{lccccc}[t!]
\label{tab:MGSDresolutions}
\centering
\tabletypesize{\small}
\tablecaption{Molecular Gas Surface Density Profiles}
\tablehead{
\colhead{Res.} & \colhead{Sample} & \colhead{16$^{th}$} & \colhead{50$^{th}$} & \colhead{84$^{th}$}}
\startdata 
60~pc & 7 SNe & 150 & 330 & 550 \\
  & All Pixels & 26 & 64 & 260 \\
\hline
90~pc & 19 SNe & 110 & 300 & 590 \\
  & All Pixels & 21 & 55 & 240 \\
\hline
120~pc & 47 SNe & 74 & 330 & 790 \\
  & All Pixels & 20 & 61 & 360 \\
\hline
150~pc & 59 SNe & 77 & 320 & 740 \\
  & All Pixels & 18 & 55 & 330 \\
\hline
500~pc & 62 SNe & 61 & 220 & 650\\
  & All Pixels & 11 & 33 & 180 \\
\hline
750~pc & 63 SNe & 58 & 210 & 550 \\
  & All Pixels & 10 & 29 & 140 \\
\hline
1000~pc & 63 SNe &  46 & 160 & 470 \\ 
  & All Pixels & 10 & 27 & 120 \\
\enddata
\tablecomments{Percentiles in units of $[{\rm M}_\odot~{\rm pc}^{-2}]$}
\end{deluxetable}

\begin{figure*}[hbtp]
    \centering
    \includegraphics[width=1.0\textwidth]{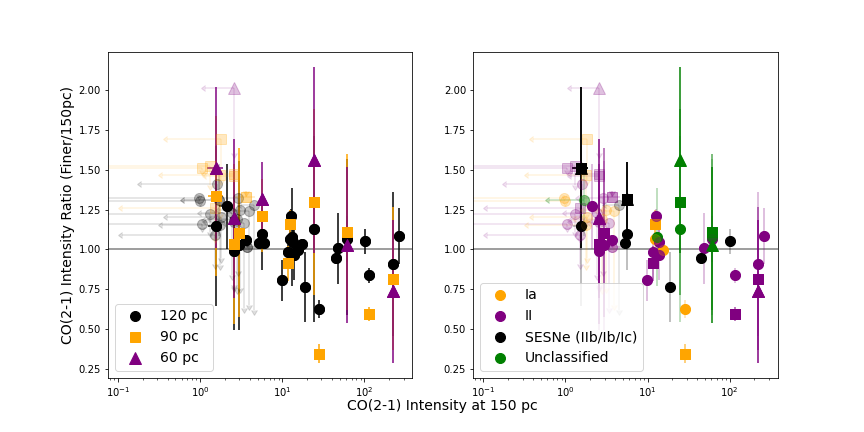}
    \caption{\textit{CO~(2-1) intensities at finer resolution vs. that at 150 pc resolution}. We plot the CO~(2-1) emission intensities as ratios of each finer resolution (120~pc in black circles, 90~pc in orange squares, and 60~pc in purple triangles) value divided by the value at 150~pc resolution on the $y$ axis and compare this to each SN location’s 150~pc CO~(2-1) intensity value on the $x$ axis (left panel). Original intensities with signal-to-noise $<$ 3 are classified as non-detections and assigned upper-limits of 3 times the noise value before taking the ratio. Non-detections are plotted with transparent symbols with error-bar-length arrows. We repeat this exercise on the right panel, now separating the SN sample by type. SNe Ia are plotted with orange markers, SNe II with purple, SESNe in black and Unclassified in  green. The marker shapes indicate the finer resolution included in the ratio, using the same marker shapes as the left panel. For both panels, the points in vertical alignment come from the same SN site. An increasing ratio at higher/finer resolution indicates that the SN is more centrally located with respect to the surrounding gas.}
    \label{fig:ResIntCompare}
\end{figure*}

In Figure \ref{fig:ResIntCompare}, we focus directly on the sites of SN explosions with measurements at multiple resolutions. In both panels, we compare the intensity of CO~(2-1) emission measured at 150~pc resolution to the ratio of the CO~(2-1) emission intensity measured at finer resolution to the 150~pc measurement for each SNe site with a 60, 90, or 120~pc resolution map. Non-detections that are assigned upper-limit values are plotted with transparent markers and with error-bar-length arrows. Points that are in vertical alignment -- that have the same $x$-axis value for each resolution -- are taken from the same SN site. By design, the upper-limit measurements (transparent markers) that have vertical alignment will have increased intensity at higher resolutions because the noise increases at resolutions finer than 150~pc. In the left panel we consider all of the SN sites, in the right panel we differentiate our sample by SN type.

For the points that have CO~(2-1) detections, the plot shows mixed results. Some of our SN sample show increased intensity at finer resolution, implying that they are directly associated with molecular gas that was partially ``smoothed out'' at coarser resolution. Others show a decrease in intensity, implying that at least a portion of the measured molecular gas coincident with the SN at 150~pc is not centrally located at the SN’s explosion site. Overall, there appears to be a trend that the SNe with lowest intensity at 150~pc resolution show at least a moderate increase in intensity at finer resolution, suggesting that the SNe are near to or within molecular clouds in these regions. 

Separating the sample by SN type we find mixed results, with all except for the unclassified population exhibiting cases of both increasing and decreasing intensity ratios at higher resolutions. The unclassified sample does show consistently higher CO~(2-1) emission intensities at higher resolutions, which perhaps reflects the difficulty in obtaining spectra for classification in high extinction regions. This plot shows that the local environments of our SNe are varied and accounting for their feedback is complicated. This also speaks to the importance of pursuing these environments with even higher resolution observations in the future.

\subsection{CO~(2-1) Distributions Around the Sites of Individual SNe}
\label{sec:zooms}

So far we have considered only emission directly at the location of the SNe, but the position of the SNe relative to nearby CO~(2-1) emission may also hold important information. As a final exercise, we explore the spatial distribution of CO~(2-1) near our SNe. To do this, in this section with construct cutout images using the highest resolution CO~(2-1) map for each SNe and in \S \ref{sec:MCdistance} we analyze the distance from each SNe to the nearest molecular gas detection.

Figure \ref{fig:Zooms} shows $500~{\rm pc} \times 500~{\rm pc}$ cutouts of each of the 59 SN explosion sites that have CO~(2-1) maps with at least 150~pc resolution. These ``zoom in" images allow us to visually locate the SN explosion sites relative to the nearby molecular gas. Inspecting the images, we explore whether the SNe with CO~(2-1) detection are located towards the peaks of the CO~(2-1) emission, or if they lie off to the side, sharing space with an MC that just happens to be caught in the beam. This latter case might arise if the SNe drifted away from their parent cloud or if other forms of feedback or previous SNe already dispersed the parent cloud. Similarly, we can inspect whether SNe with non-detections might have nearby molecular gas that is just not directly coincident with the SNe.

Examining the maps, we note several impressions. First, we find 100\% agreement between the detection classification at 150~pc and at those at finer resolutions from (60-120~pc). Second, the cutouts show that the molecular gas around the SNe is often extended at the resolution shown and sometimes patchy. Non-detections often lie in the vicinity of CO~(2-1) emission. Detections are often not on the emission peaks themselves but off to the side. Overall, the zooms show that the SNe are exploding into environments that have a wide range of molecular gas densities and landscapes and tend to show a high degree of local variation in density.

We study recent SNe that have not yet affected their environments. We plot three “spheres of future influence” in order to estimate the size of the region they will affect in the future and so form an idea of how these explosions may change the nearby ISM. White circles show apertures with radii of 6.3, 32, and 203~pc around each SNe site. The 6.3~pc aperture is an estimate of the cooling radius\footnote{The cooling radius is the spatial scale over which the ISM receives a significant momentum boost from the SN explosion (i.e. momentum feedback), as confirmed by recent simulations \citep[e.g.,][]{Martizzi2015,Kim2015}. During this time, radiative losses become important during the evolution of the SNR.} of an SNR in the turbulent and inhomogeneous ISM at a typical molecular gas density ($n_{H} = 100 {\rm cm}^{-3}$) \citep{Martizzi2015}. The 32 and 203~pc apertures represent the cooling and fadeaway\footnote{The fadeaway radius is the physical point where the shock front of the SNR slows to match the surrounding ISM's effective sound speed.} radii for a SNR in low-density ($n_{H} = 0.5 {\rm cm}^{-3}$) environments \citep{Draine2011}. Although the actual radii are dependent on the exact density structure of the ISM and the energy of the SN, these estimates allow us to visualize how much of the surrounding ISM could be influenced by the SNe that have just exploded.

Inspecting the individual SN sites in Figure \ref{fig:Zooms} we identify five general cases:

\begin{enumerate}
\item \textit{Non-detections with CO~(2-1) within the cooling radius.} A few of the CO~(2-1) non-detections show nearby emission that suggests future interaction between the SN and the nearby molecular gas. Specifically, 7\% (4/59) of the panels show the presence of molecular gas within the low-density cooling radius of a non-detection classification. Specific SNe: SN1935C, SN2016cok, SN1964F, \& SN2007it.

\item \textit{Non-detections with CO~(2-1) within the fadeaway radius.} An additional 24\% (14/59) show molecular gas within the fadeaway radius of a non-detection classification. We expect these SN to be likely to interact with the surrounding ISM sometime before fading, though the interaction may be relatively weaker than for the first group of non-detections. Specific SNe: SN1985P, SN2006ov, SN2014dt, SN2006X, SN2019ehk, SN1960F, SN1981B, SN2008bk, SN2013ej, SN2001du, SN1986L, SN1993Z, SN2020jfo, \& SN1979C.

\item \textit{Detections with nearby voids or low-density regions.} 8\% (5/59) of our sample show CO~(2-1) detections but also the presence of nearby voids or lower-density regions on the edge or just outside the high-density cooling radius at our detection classifications. We expect the explosion blast wave to move more quickly through these lower-density regions, while slowing through interaction with the denser ISM in the higher-density areas. Specific SNe: SN2009ib, SN1972Q, SN1923A, SN1983N, \& SN2005at.

\item \textit{Detections in regions with widespread high surface densities.} 47\% (28/59) occur in apparently dense regions where CO~(2-1) emission extends completely past the high-density cooling radius. These are the most likely cases for the the energy of the SN to contribute to the destruction or reshaping of the nearby MCs without breaking through to the surrounding low-density ISM. We caution however that the three-dimensional distribution and substructure below our resolution remain unknown, so even these cases may still have low-density channels through which some of the explosion can exert a more extended effect. Specific SNe: SN1996cr, SN1940E, SN1995V, SN1992bd, SN1983V, SN1984J, ASASSN-14ha, SN2010el, SN1999em, SN2003jg, SN2008eh, SN1973R, SN1989B, SN2009hd, SN1967H, SN1986I, SN2014L, SN1999gn, SN1901B, SN1959E, SN2020oi, SN2020nvb, SN2011ja, SN1986G, SN2016adj, SN1950B, SN1957D, \& SN1968L.

\item \textit{Non-detections with little or no nearby molecular gas.} The remaining 14\% (8/59) occur isolated from nearby detected molecular gas extending all the way past the low-density fadeaway radius. These SNe are the most likely ones to drive turbulence in the atomic gas, carve out large bubbles and shells, potentially launch outflows, and otherwise interact primarily with the diffuse ISM. Specific SNe: SN1999eu, SN2012fr, SN2005df, SN2012A, SN1926A, SN1961I, SN2012cg, \& SN1988M.
\end{enumerate}

\begin{figure*}[!h]
    \centering 
    \subfloat[]{%
        \includegraphics[width=0.95\linewidth]{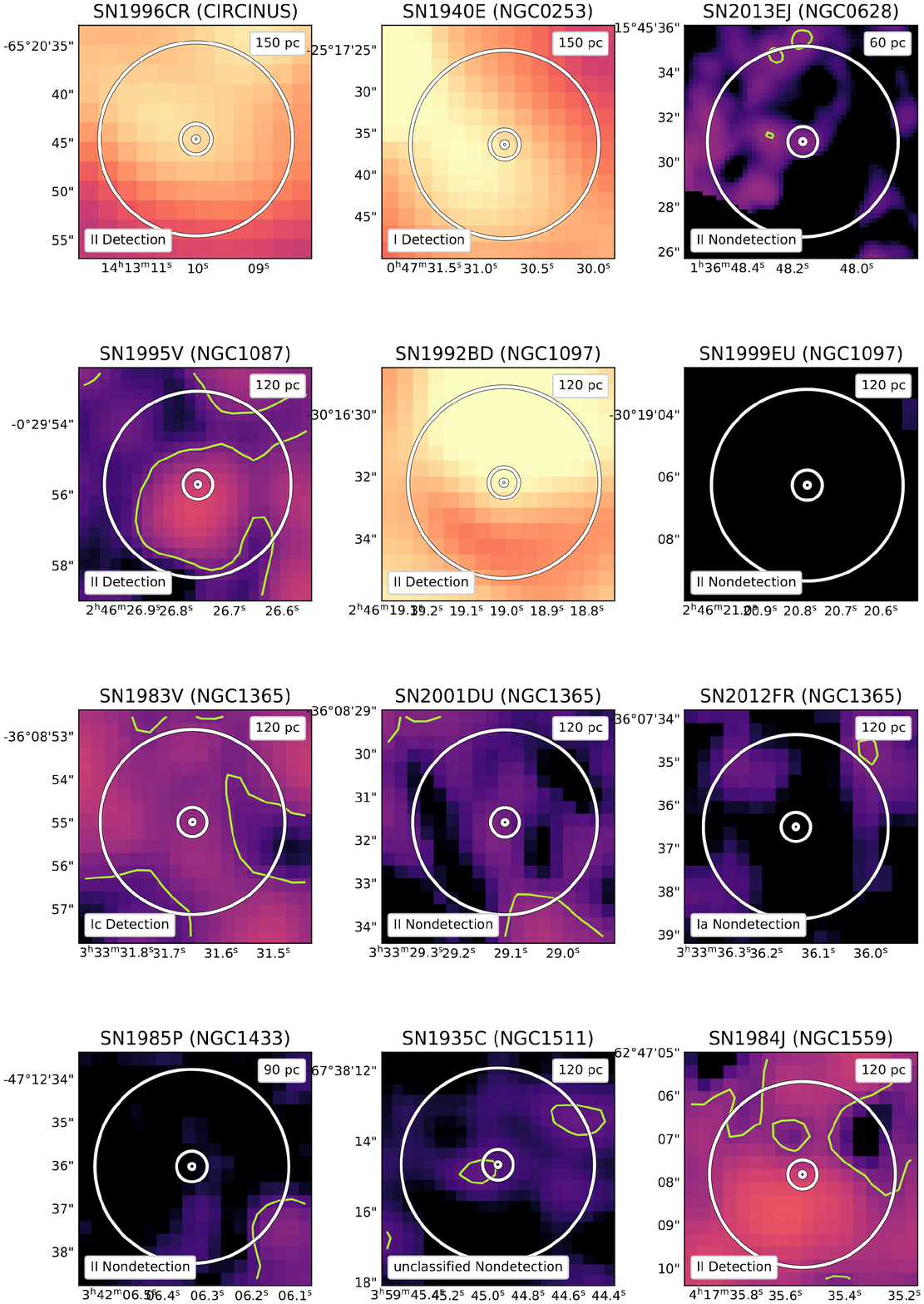}
        \label{fig:6.a}%
    }
\caption{\textit{CO~(2-1) emission cut-outs ($500 \times 500~pc$) centered on the locations of 59 SNe that have PHANGS--ALMA coverage from 60-150~pc.} The spatial resolution of each cutout is marked in the top right corner of each panel. Spheres of future influence are plotted with radii at 6.3 (cooling radius for an inhomogeneous, typical-density, $n_H = 100 cm^{-3}$, ISM), 32 \& 203~pc (cooling and fadeaway radii for a low-density, $n_H = 0.5 cm^{-3}$, ISM). Lime contours enclose CO~(2-1) emission with signal-to-noise $\geq3$. Each SN is labeled with its detection/non-detection assignment given from its signal-to-noise measurement at 150~pc resolution.}
\end{figure*}
\begin{figure*}[!h]\ContinuedFloat                          
\subfloat[]{%
        \includegraphics[width=0.95\linewidth]{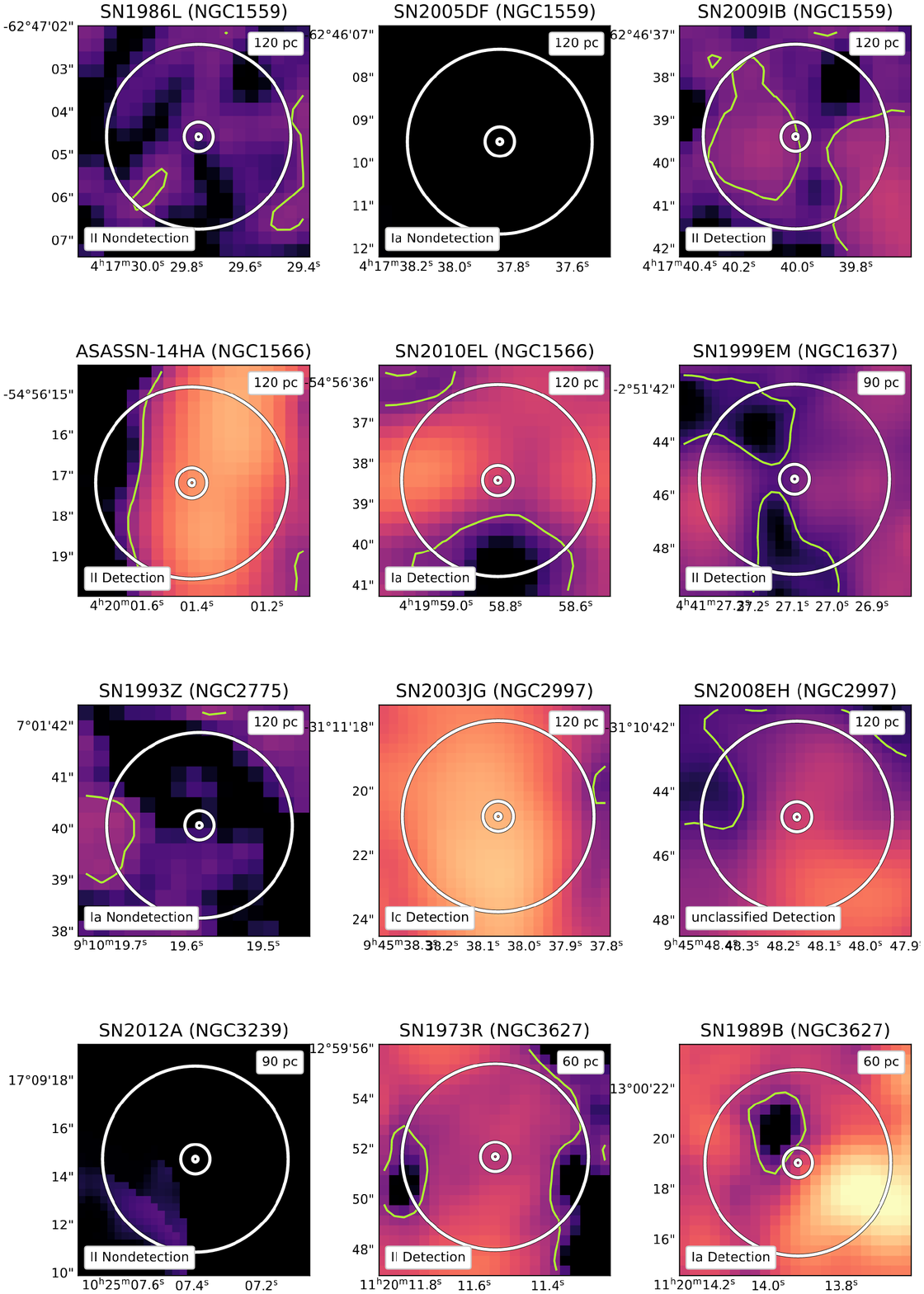}
        \label{fig:6.b}%
        }
\caption{Continued}
\end{figure*}
\begin{figure*}[!h]\ContinuedFloat                          
\subfloat[]{%
        \includegraphics[width=0.95\linewidth]{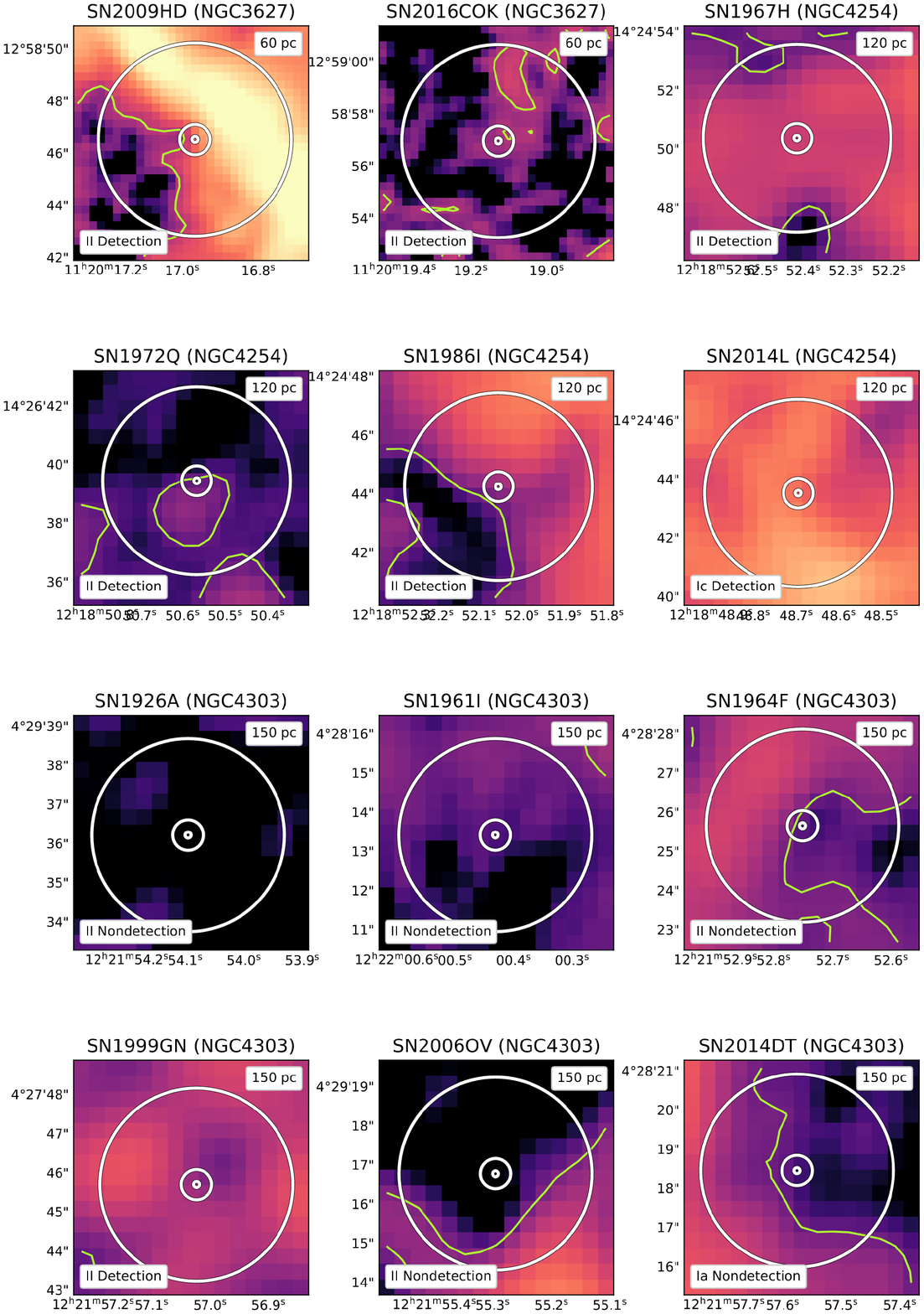}
        \label{fig:6.c}%
        }
\caption{Continued}
\end{figure*}
\begin{figure*}[h!]\ContinuedFloat                          
\subfloat[]{%
        \includegraphics[width=0.95\linewidth]{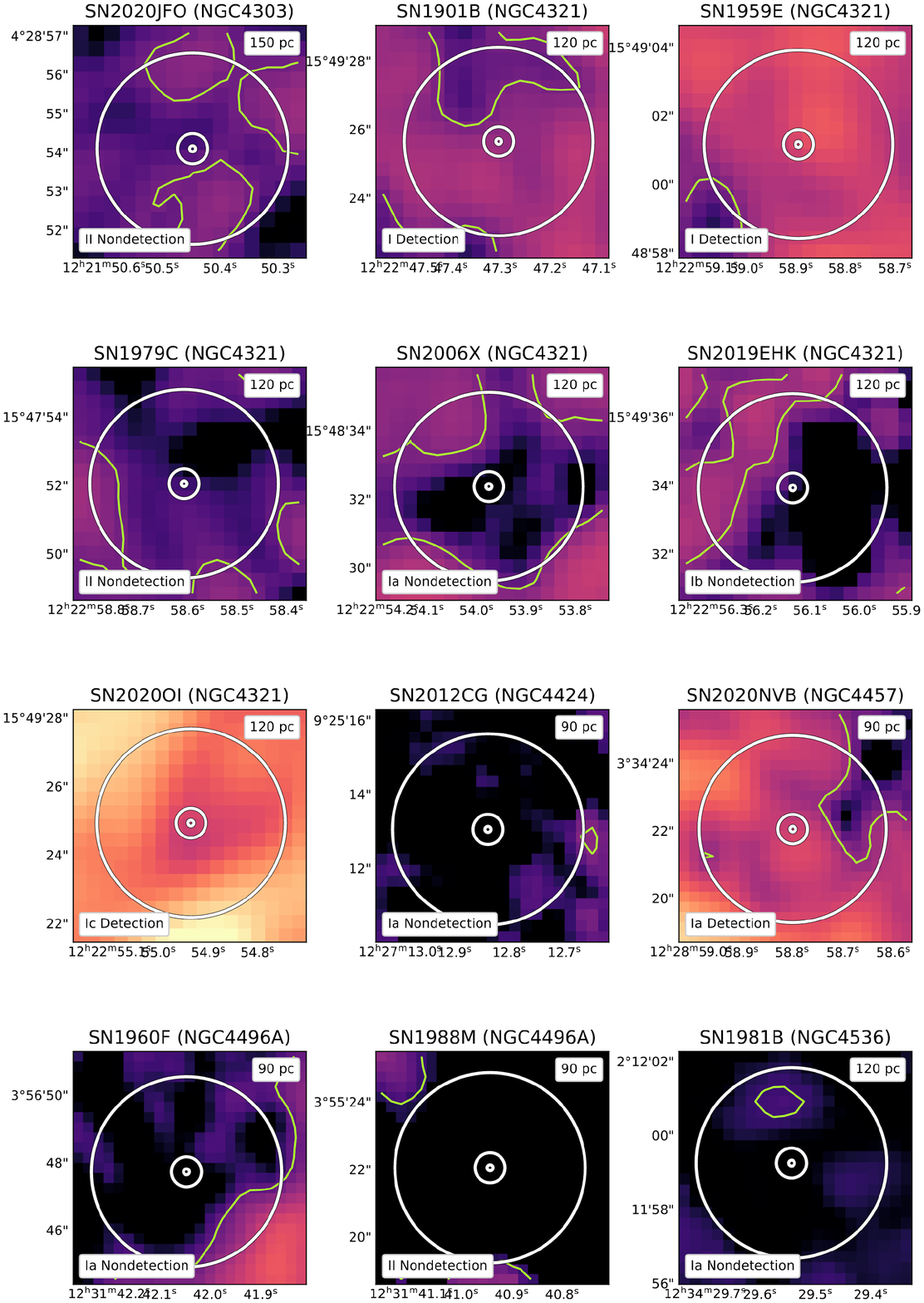}
        \label{fig:6.d}%
        }
\caption{Continued}
\end{figure*}
\begin{figure*}[!h]\ContinuedFloat                          
\subfloat[]{%
        \includegraphics[width=0.95\linewidth]{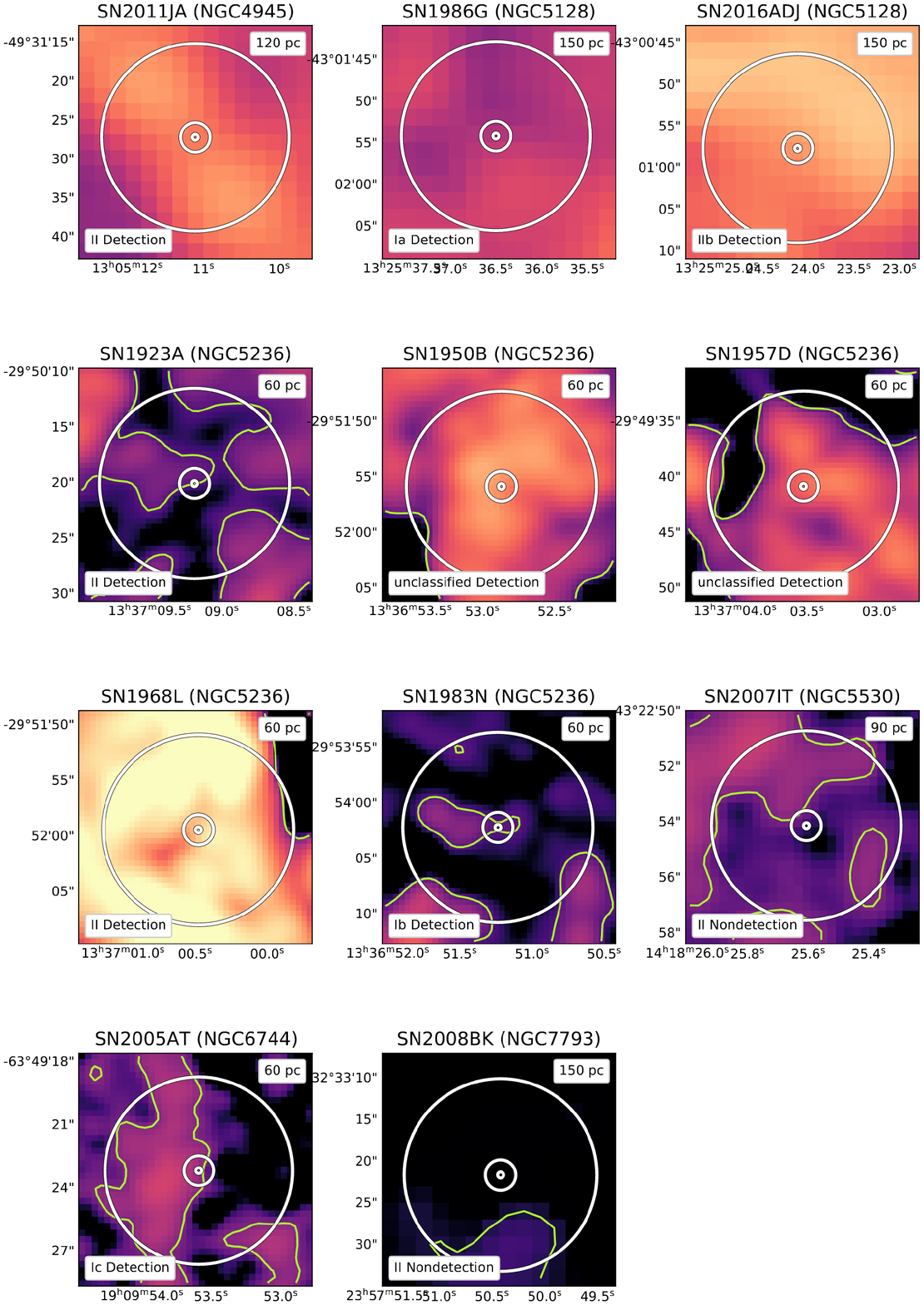}
        \label{fig:6.e}%
        }
\caption{Continued}
\label{fig:Zooms}
\end{figure*}


\begin{deluxetable*}{lcccccc}[t!]
\label{tab:cutoutCases}
\centering
\tabletypesize{\small}
\tablecaption{SN Type Fractional Distribution of Cutout Cases}
\tablehead{
\colhead{Case} & \colhead{Total SNe}& \colhead{SN Ia} & \colhead{SN II} & \colhead{SESNe} & \colhead{Unclassified}}
\startdata 
1. Non-detections with SNe within cooling radius & 4 & 0 & 3  & 0 & 1 \\
\hline
2. Non-detections with SNe within fadeaway radius & 14 & 5 & 8 & 1 & 0 \\
\hline
3. Detections with nearby voids & 5 & 0 & 3 & 2 & 0 \\
\hline
4. Detections in widespread high surface density & 28 & 4 & 14 & 4 & 6 \\
\hline
5. Non-detections with no nearby CO & 8 & 3 & 5 & 0 & 0\\
\enddata
\end{deluxetable*}

Table \ref{tab:cutoutCases} shows the SN type distribution among the five different cutout cases described above. Cases 1, 2, and 5 represent environments with the lowest amounts of nearby molecular gas. Case 1 is predominantly SNe II while Cases 2 and 5 have the highest fraction of SNe Ia compared to the other cases. In Case 1 we find predominantly SNe II that do not have strong enough CO~(2-1) emission to be considered a detection, yet they have molecular gas coincident within $\sim30$ parsecs. This is consistent with CCSNe progenitors with lifetimes on the order of $\sim10$s of Myrs that will have time to migrate some distance away from their parent cloud before exploding. Cases 2 and 5 represent the SN environments most devoid of molecular gas and host the highest fractions of SNe Ia and SNe II. Collectively, Cases 1, 2, and 5 support the weakest association between molecular gas and SNe Ia and II. Cases 3 and 4 host 86\% (6/7) of the total SESNe and unclassified SNe populations and clearly show that these are more associated with the highest molecular gas surface density regions. This is consistent with SESNe originating from some of the most massive progenitors and having the shortest lifetimes.

Even more than the statistical analysis, the cutouts clearly show that even in gas-rich regions of massive galaxies, SNe must play a varied role in exerting feedback on their surrounding environment, occurring across a wide range of molecular gas densities. Given the patchy gas distribution and the position of the SNe near both high- and low-column density regions, most of our SNe (even many of our non-detections) may both plausibly interact with molecular gas in the future \textit{and} have a channel to affect the lower-density, diffuse ISM. It seems likely that individual SNe, many of which lie near the edges of molecular structures, play multiple roles, both affecting the dense and diffuse ISM.


\subsection{Distance from SNe to the nearest detected molecular gas}
\label{sec:MCdistance}

Finally, as a quantitative complement to \S \ref{sec:zooms}, we measure the distance from each SNe to the nearest pixel with well-detected molecular gas at each galaxy's individual median $3\sigma$ $\Sigma_{\rm mol}$ value. These $3\sigma$ values range from 3.6 to 17$ M_\odot~{\rm pc}^{-2}$ with a median value of 7.9$ M_\odot~{\rm pc}^{-2}$. We compare these measured distances for our real SN sample to four model populations, each constructed to reflect a distinct hypothesis for where SNe explode in our targets. Figure \ref{fig:Models} shows how we generate these model SNe populations. The four hypotheses are: 

\begin{figure*}[h!]
    \centering
    \includegraphics[width=1.0\textwidth]{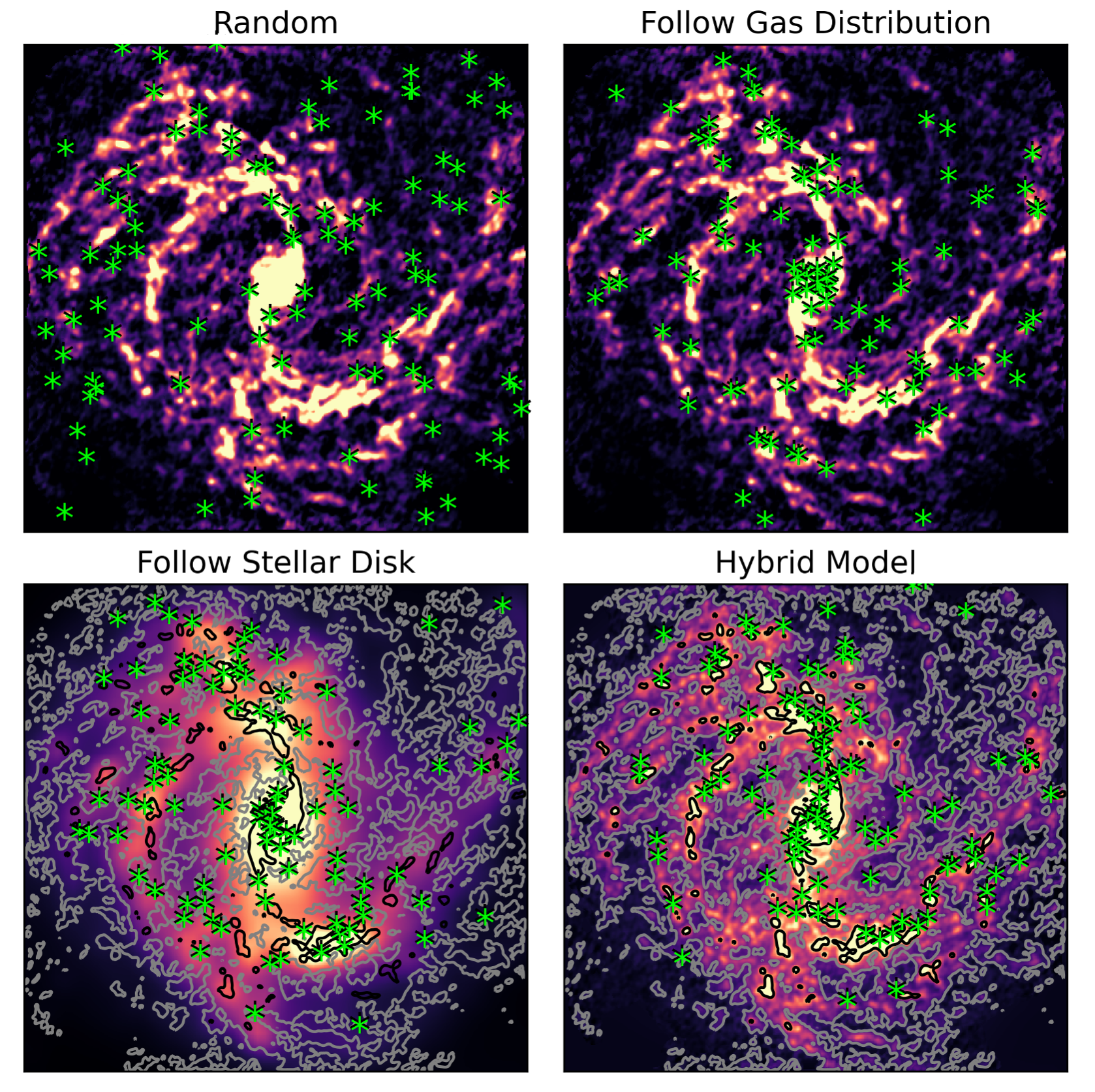}
    \caption{\textit{Four models for 100 generated SNe within galaxy NGC4303.} From left to right -- Model 1: SNe randomly generated across the footprint of PHANGS--ALMA CO~(2-1) map. Model 2: SNe generated from a probability distribution that follows the gas distribution of the CO~(2-1) map. Model 3: SNe generated from a probability distribution that follows the stellar population from 3.6 $\mu$m map from the Infrared Array Camera (IRAC) on \textit{Spitzer}. Model 4: 76\% percent of SNe are pulled from the gas distribution while the remaining 24\% are pulled from the stellar distribution. Underlying map image is a combined Infrared + CO~(2-1) emission map using the model's flux ratio. Grey and black contours mark regions with signal-to-noise $\geq3$ for CO~(2-1) and IR emission respectively.}
    \label{fig:Models}
\end{figure*}

\begin{enumerate}
\item \textit{Purely random:} the SN distribution is completely random with SNe equally likely to occur in each pixel of the map. This would be a fitting model if SNe simply occurred randomly in space with no relation to their parent galaxy. This is not expected but represents the simplest control, so we calculate it.

\item \textit{Following the molecular gas distribution:} the SN distribution favors pixels at local peaks in the density field. We assign a probability that a SN will occur at a given pixel based on the intensity profile of the PHANGS--ALMA CO~(2-1)  maps. This would be expected if the gas distribution is long-lived in its current configuration, the amount of gas determines the amount of star formation, and the SNe overall trace the locations of star formation. Because CCSNe trace high-mass star formation and high-mass star formation follows the molecular gas, we expect this to be a good general model for the placement of CCSNe. This model will be biased to preferentially place SNe in galaxy centers, although, observationally, SN are difficult to observe in these dense, bright regions, leaving our real SN sample with fewer SNe from these regions. 

\item \textit{Following the stellar disk:} the SN distribution follows the distribution of stellar mass in the galaxy, which tends to be smoother than the gas distribution and overall resembles an exponential disk in most targets. We assign a probability that a SN will occur at a given pixel based on the intensity profile of the near-IR maps described in \S\ref{sec:Methods}. Type Ia SNe have been shown to follow the stellar disk \citep{Maoz2014, Anderson2015a, Cronin2021}. For this reason, we would expect this model to provide good placement for Type Ia SNe.

\item \textit{Hybrid model:} We combine the two previous models, generating 24\% of our SN population to occur following the stellar disk and 76\% following the gas distribution. This ratio is set from previous work that has shown that for massive star forming galaxies 24\% of observed SNe are Type Ia \citep{Li2011a}, which closely resembles our SNe population of 23\% SNe Ia. 
\end{enumerate}

We generate 100 model SNe for each galaxy in our sample for which we have a map at 150~pc resolution\footnote{missing 150 pc CO~(2-1) maps for NGC 1068, 1672, and 4579}. This gives us a total of 2900 model SNe from 29 galaxies generated for each of the four models.

\begin{figure}[]
    \centering
    \includegraphics[width=0.5\textwidth]{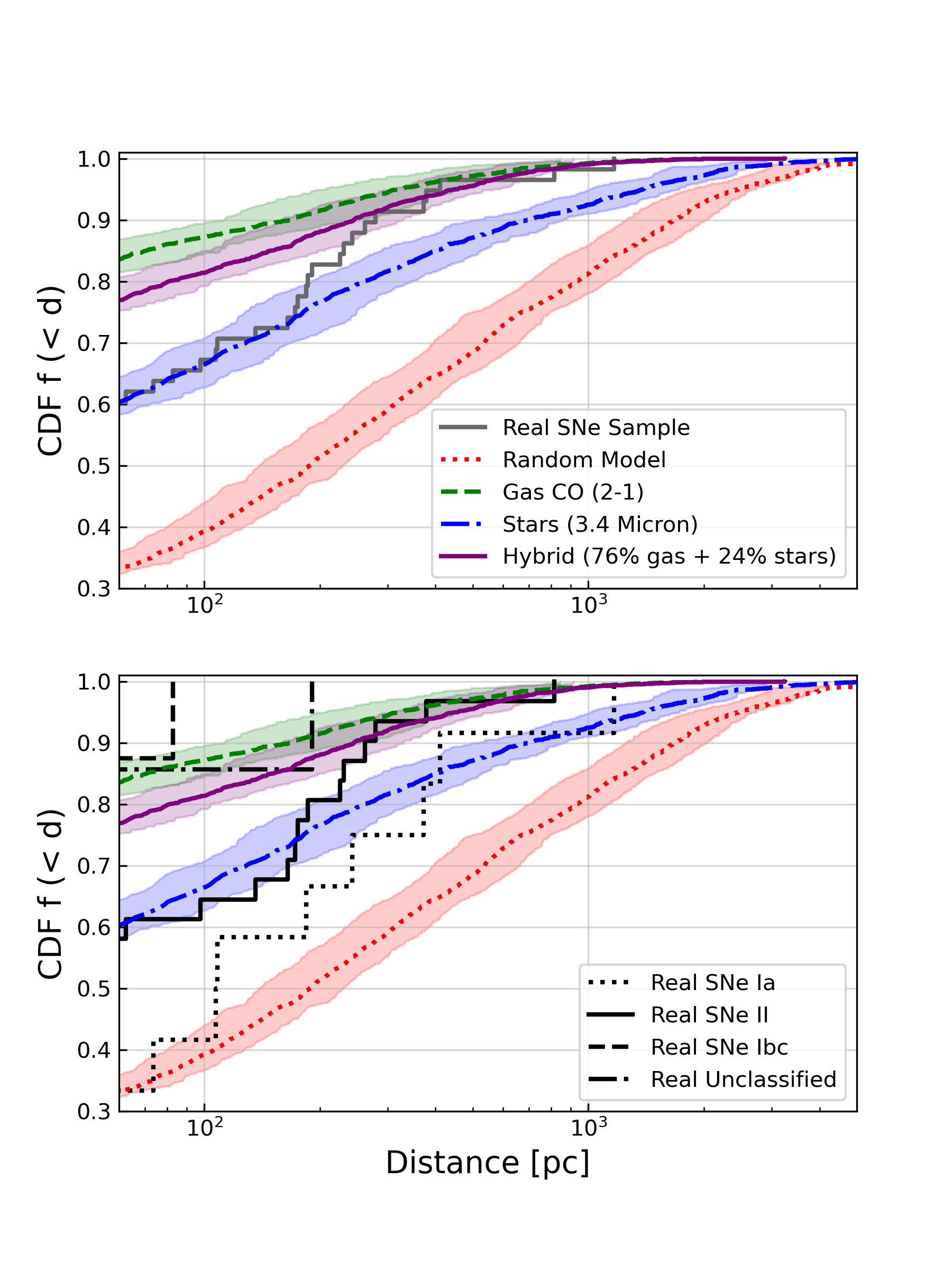}
    \caption{\textit{CDFs of distances to the nearest pixel with molecular gas surface density of each galaxy's individual median $3\sigma$ $\Sigma_{\rm mol}$ or higher}. Our real SNe sample is drawn with a dark-grey line. The randomly generated SNe sample is drawn with a dotted red line, gas distribution as dashed green line, stellar as dash-dotted blue line, and the hybrid model as solid purple line. The transparent shading represents the 16th-84th percentile values from 1000 random pulls each the size of our observed SN sample from each model distribution. In the bottom panel, we separate our observed SN sample by SN type. SNe Ia are marked with a dotted-black line, SNe II with a solid line, SNe Ibc with a dashed line and unclassified SNe with a dashed-dotted line. The models are plotted with the same color scheme as the top panel.}
    \label{fig:modelCDFs}
\end{figure}

In Figure \ref{fig:modelCDFs} we show the cumulative distributions of the distances to the nearest pixel with each galaxy's individual median $3\sigma$ $\Sigma_{\rm mol}$ value for both our real SN sample (59 SNe) and each of our four model SN populations (2900 SNe generated for each model). In the top panel we plot our real SN sample together regardless of type, in the bottom panel we separate our real SN sample by type. In the top panel of Figure \ref{fig:modelCDFs}, the random model produces very unrealistic results, with little or no association between the SNe and CO~(2-1) emission. On the other hand, the distribution following gas directly produces too close of an association. The hybrid model does significantly better, with the observations producing a distance distribution between the stellar disk model at small distances and the hybrid model at larger distances. We find that almost $\sim65\%$ of our SN sample is within 100 pc of a molecular gas surface density of our $3 \sigma$ value of $7.9 M_\odot {\rm pc}^{-2}$ or higher. In the bottom panel of Figure \ref{fig:modelCDFs} we find that the SESNe and untyped SNe are found close to well-detected molecular gas, with 100\% of our SNe Ibc and 85\% of our untyped SNe occurring with a CO~(2-1) detection within the beam at 150~pc. Approximately 60\% of our SNe II occur with CO~(2-1) detection in the beam with $\sim90\%$ occurring within 300~pc of molecular gas. SNe Ia tend to be found at a wider range of distances to molecular gas with $\sim60\%$ occurring with a CO~(2-1) detection in the beam with $\sim90\%$ occurring within 400~pc of molecular gas.

\begin{figure}[]
    \centering
    \includegraphics[width=0.5\textwidth]{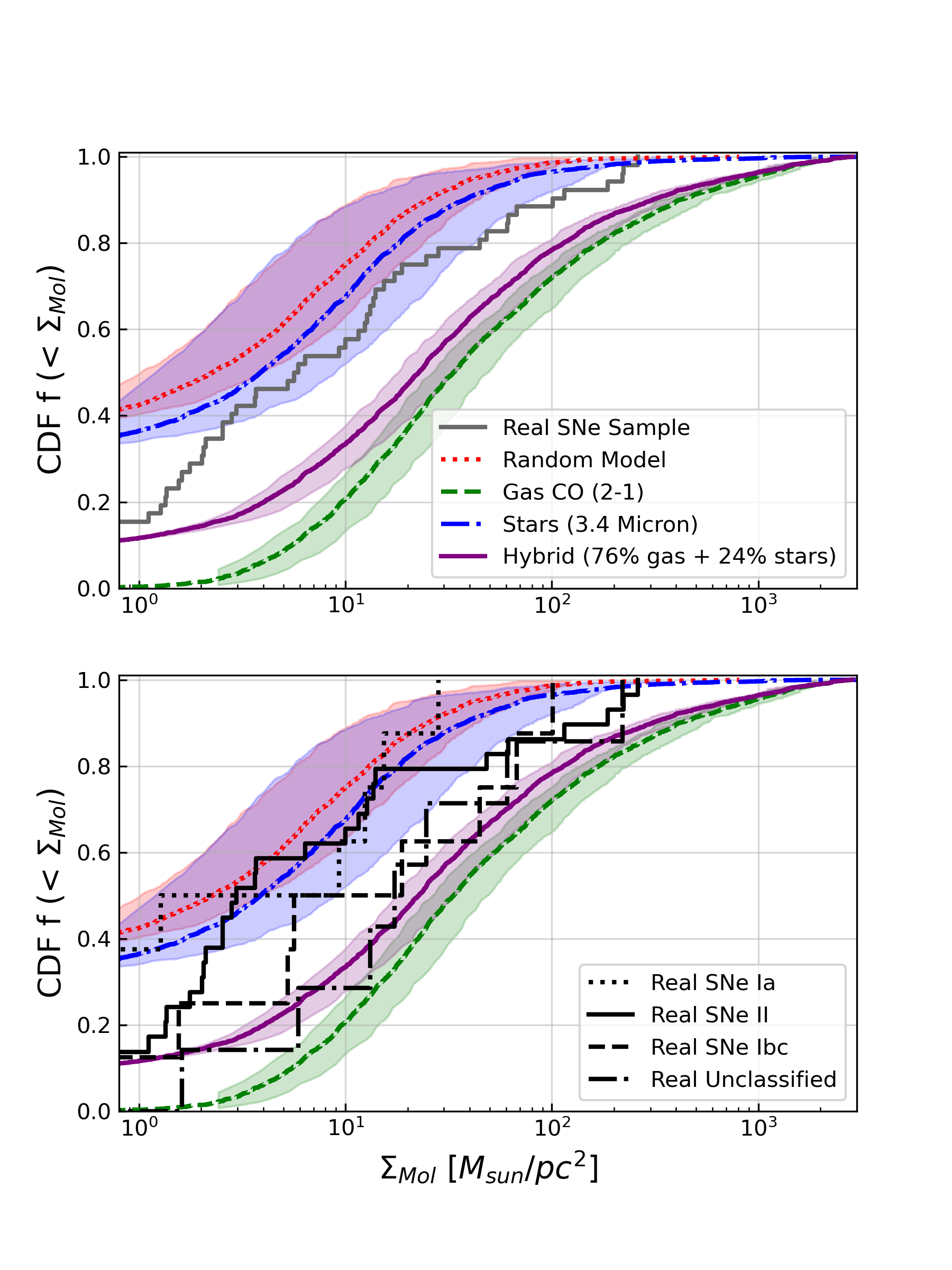}
    \caption{\textit{CDFs of mass as a function of the molecular gas surface density at the locations of our SN sample at 150~pc resolution compared to our model SN}. In the top panel we compare our observed SNe sample with each of our models. Our real SNe sample is drawn with a dark-grey line. The randomly generated SNe sample is drawn with a dotted red line, gas distribution a dashed green line, stellar a dash-dotted blue line, and the hybrid model a solid purple line. The transparent shading behind each model line represents the 16th-84th percentile values from 1000 random pulls each the size of our observed SN sample from each model distribution. In the bottom panel, we separate our observed SN sample by SN type. SNe Ia are marked with a dotted-black line, SNe II with a solid line, SNe Ibc with a dashed line and unclassified SNe with a dashed-dotted line. The models are plotted with the same color scheme as the top panel.}
    \label{fig:modelCDFsSM}
\end{figure}

Figure \ref{fig:modelCDFsSM} is constructed similarly to Figure \ref{fig:modelCDFs} but instead we show the cumulative distributions of the mass as a function of the molecular gas surface density at the locations of our SN sample at 150~pc resolution for both our real SNe sample (59 SNe) and each of our four model SNe populations (2900 SNe in each model). We find that the CDF of our observed SNe sample lies between the stellar disk model and the hybrid model, but more closely follows the stellar disk model. In the bottom panel we separate the observed SNe by type and repeat the comparison. We find the SNe Ia and II more closely follow the random and stellar disk model respectively, the SNe Ibc more closely follow between the stellar disk model and the hybrid model, and the unclassified SNe more closely follow the hybrid model.

\section{Discussion and Conclusions}
\label{sec:Discussion}
 
We perform the first statistical characterization of the molecular gas content at the location of SN explosion sites on $\sim100$~pc scales. This is the first time an observational study on feedback from individual SNe has been done on the scales of individual GMCs. 

The effects of SNe on their host galaxies are location dependent. If a SN explodes within an MC, its energy and momentum may work to disperse that cloud. On larger scales, SN explosions may collect and compress gas, helping to trigger star formation. And if a SN explodes in the diffuse ISM, the explosion may be well-placed to create turbulence in diffuse gas or launch galactic winds or fountains. 

Until recently, it was not possible to measure the gas content of SN explosion sites at resolutions high enough to associate individual explosions with gas likely to interact with that SN. Thanks to large samples of SN discoveries and new high-resolution CO~(2-1) surveys with ALMA, this situation is changing. 

We measure the CO~(2-1) emission from the sites of $>60$ recent ($\lesssim 100$ years) SNe, which allows us to place SN explosions in the context of their host galaxies and investigate the role that feedback from these explosions could play in the evolution of GMCs.

\begin{enumerate}

\item We identified 63 SNe from the OSC that exploded in 31 galaxies from the PHANGS-ALMA footprint (see Figure \ref{fig:AllGalaxies} and Table \ref{tab:OSCSample}). We measure the CO~(2-1) intensity at the SN locations across a wide range of resolutions from 60~pc to 1~kpc (see Table \ref{tab:intensities}). We find that $\sim95\%$ of our SN explosion sites have detectable CO~(2-1) emission on 1~kpc scales and $\sim60\%$ on 150~pc scales (see \S \ref{sec:detectionallres}, Figure \ref{fig:SNRAcrossResolutions}: Table \ref{tab:SampleAcrossResolutions}, and Table \ref{tab:MGSDresolutions}). Primarily this difference between results at different scales appears to reflect that SNe explode in gas-rich regions of galaxies but that they do not always appear directly coincident with gas peaks at higher resolution. This is reinforced by Figure \ref{fig:Zooms}, where we zoom-in to the immediate environment of each of our SNe and find that most of them occur off-peak of the nearby molecular gas.

\item We construct a working sample of 59 SNe that have CO~(2-1) intensity measurements at 150pc resolution (see \S \ref{sec:detection150pc}). We compare distributions of molecular gas surface densities at the sites of these SN explosions to the overall distributions in the PHANGS–ALMA CO~(2-1) maps. We find that SNe consistently explode in regions with higher than average molecular gas surface densities (see Figure \ref{fig:CDF_ALLSNe}, Figure \ref{fig:CDFByType}, and Table \ref{tab:150MGSDAverages}). We find this despite the fact that one might expect optical SN searches to be biased against SN that occur in the most gas-rich parts of galaxies due to the effects of extinction.

\item When separating our SN sample by type, we find that our CCSNe (SNe II and SESNe -- IIb/Ib/Ic) occur in brighter CO~(2-1) emitting regions than our SNe Ia (see Figure \ref{fig:CDFByType} and Table \ref{tab:150MGSDAverages}). Our sample of SESNe show the strongest association with bright CO~(2-1) emission, and have the highest fraction of CO~(2-1) detections of all SN types occurring in the galaxies' highest density regions. In turn, SNe II have a stronger association with CO~(2-1) emission than SNe Ia. Taking these results together, we find an SN's association with peaks in molecular gas density to decrease as a function of progenitor mass (see Figure \ref{fig:CDFAllTypes}). We find that the unclassified SNe are most consistent with the SESNe population (see Figure \ref{fig:CDFByType} and Figure \ref{fig:modelCDFsSM}), with $\sim75\%$ of the unclassified SNe coincident with dense GMCs (see Figure \ref{fig:Zooms}). It is possible that spectroscopic follow-ups of transients could miss SESNe in regions with high extinction.

\item We plot CO~(2-1) emission in a local $500~{\rm pc} \times 500~{\rm pc}$ region centered on each SN in our sample in order to visually inspect the SN location (see Figure \ref{fig:Zooms}). The images show that even when CO~(2-1) is detected SNe often occur displaced from the local peak of CO~(2-1) emission, and the surrounding region often shows a wide range of molecular gas surface densities. Meanwhile, a significant fraction of SNe with non-detections have some nearby CO~(2-1) that might plausibly interact with the SN explosion before it fades away.

\item Finally, we model four different populations of SN locations and compare their distributions with that of our observations (see Figure \ref{fig:Models}). We find that CDFs of both the distance to the nearest well-detected molecular gas and the surrounding molecular gas surface density of our observed SN sample falls between the locations of modeled SNe generated to follow the infrared emission of older stellar populations and the locations of SNe generated to follow a hybrid model with $\sim75\%$ molecular gas and $\sim25\%$ stellar infrared emission. (see \S \ref{sec:MCdistance}: Figure \ref{fig:modelCDFs} and Figure \ref{fig:modelCDFsSM}).

\end{enumerate}

These results show that the molecular gas landscape in the vicinity of recent SNe is varied, and we can expect that the impact of a SN on its local environment has the potential to be complex. This speaks to the importance of having high-resolution measurements to better describe these environments in the future.

We expect that both the SN sample and the gas data used for this study will improve over the next few years. We use a SN sample drawn from heterogeneous observations over the last century, and these observations often have unknown completeness and biases against finding SNe in dusty, high-extinction regions like MCs or galaxy centers. 

Systematic surveys that image the entire sky at high cadence -- such as the All Sky Automated Search for SN \citep[ASAS-SN,][]{Shappee2014, Kochanek2017} and the Vera C. Rubin Observatory’s Legacy Survey of Space and Time \citep[LSST,][]{Cook2004} are changing this. Over time we expect large highly complete samples of SNe to emerge for nearby galaxies. Future SN environment studies will benefit from these surveys.

Although this work provides us an up-close view of a SN environment on a 100~pc scale, further advances will come from exploring these regions on even smaller, sub-MC, scales: CO~(2-1) observations can plausibly be pushed to resolutions of $\sim 5{-}20$~pc, enough to cleanly locate the SN relative to any MCs and resolve the future cooling radius or region of likely interaction. There are also gains to be made by improving the sensitivity. Our point-mass sensitivity for much of the survey corresponds to the mass of an individual GMC, but we cannot rule out the presence of lower-mass clouds. With deeper integration we could ascertain whether or not a lower-mass cloud is present.

Future directions include looking at CO~(2-1) emission line widths to explore if the explosions are well-placed to add turbulence to diffuse gas or to launch galactic winds or fountains, as well as looking at H$\alpha$ emission to see how often the SNe visibly reside in regions cleared by pre-SN feedback.

\acknowledgements 

NMC thanks the Ohio State University's Galaxy and Supernova groups, including Christopher Kochanek, Kris Stanek, and Patrick Vallely, for useful discussions at several stages of the project. This work was carried out as part of the PHANGS collaboration.

Support for this work was provided by the NSF through award SOSP SOSPADA-010 from the NRAO, which supported the work of NMC. The work of NMC, AKL, and JS was partially supported by the National Science Foundation (NSF) under Grants No.~1615105, 1615109, and 1653300.

The work of JS and AKL is partially supported by the National Aeronautics and Space Administration (NASA) under ADAP grants NNX16AF48G and NNX17AF39G.

The work of JS is partially supported by the Natural Sciences and Engineering Research Council of Canada (NSERC) through the Canadian Institute for Theoretical Astrophysics (CITA) National Fellowship.

MC gratefully acknowledges funding from the Deutsche Forschungsgemeinschaft (DFG, German Research Foundation) through an Emmy Noether Research Group (grant number CH2137/1-1). COOL Research DAO is a Decentralized Autonomous Organization supporting research in astrophysics aimed at uncovering our cosmic origins.

MC and JMDK gratefully acknowledge funding from the DFG through an Emmy Noether Research Group (grant number KR4801/1-1) and the DFG Sachbeihilfe (grant number KR4801/2-1), as well as from the European Research Council (ERC) under the European Union's Horizon 2020 research and innovation programme via the ERC Starting Grant MUSTANG (grant agreement number 714907).

SCOG acknowledges financial support from the DFG via the collaborative research center (SFB 881, Project-ID 138713538) ``The Milky Way System''  (subprojects A1, B1, B2, B8, and P2). They also acknowledge funding from the Heidelberg Cluster of Excellence ``STRUCTURES'' in the framework of Germany’s Excellence Strategy (grant EXC-2181/1, Project-ID 390900948) and from the European Research Council via the ERC Synergy Grant ``ECOGAL'' (grant 855130).

KG is supported by the Australian Research Council through the Discovery Early Career Researcher Award (DECRA) Fellowship DE220100766 funded by the Australian Government.

ES and TGW acknowledge funding from the European Research Council (ERC) under the European Union’s Horizon 2020 research and innovation programme (grant agreement No. 694343).

HAP acknowledges support by the Ministry of Science and Technology of Taiwan under grant 110-2112-M-032-020-MY3.

GAB acknowledges support from the ANID BASAL FB210003 project.

This paper makes use of the following ALMA data, which have been processed as part of the PHANGS--ALMA CO~(2-1) survey: \\
\noindent 
ADS/JAO.ALMA\#2012.1.00650.S\\ 
ADS/JAO.ALMA\#2013.1.00803.S\\
ADS/JAO.ALMA\#2013.1.01161.S\\
ADS/JAO.ALMA\#2015.1.00121.S\\
ADS/JAO.ALMA\#2015.1.00782.S\\
ADS/JAO.ALMA\#2015.1.00925.S\\
ADS/JAO.ALMA\#2015.1.00956.S\\
ADS/JAO.ALMA\#2016.1.00386.S\\
ADS/JAO.ALMA\#2017.1.00392.S\\
ADS/JAO.ALMA\#2017.1.00766.S\\
ADS/JAO.ALMA\#2017.1.00886.L\\
ADS/JAO.ALMA\#2018.1.00484.S\\
ADS/JAO.ALMA\#2018.1.01321.S\\
ADS/JAO.ALMA\#2018.1.01651.S\\
ADS/JAO.ALMA\#2018.A.00062.S\\
ADS/JAO.ALMA\#2019.1.01235.S\\
ADS/JAO.ALMA\#2019.2.00129.S\\
ALMA is a partnership of ESO (representing its member states), NSF (USA), and NINS (Japan), together with NRC (Canada), NSC and ASIAA (Taiwan), and KASI (Republic of Korea), in cooperation with the Republic of Chile. The Joint ALMA Observatory is operated by ESO, AUI/NRAO, and NAOJ. The National Radio Astronomy Observatory (NRAO) is a facility of NSF operated under cooperative agreement by Associated Universities, Inc (AUI).

This work acknowledges the Open Supernova Catalog (OSC; \citealt{Guillochon2017}), last accessed on 2021 May 25, which included observations and metadata for $\sim$80,000 SNe \& SNRs.

This work has made use of SAO/NASA Astrophysics Data System\footnote{\url{http://www.adsabs.harvard.edu}}, the NASA/IPAC Extragalactic Database (NED), which is operated by the Jet Propulsion Laboratory, California Institute of Technology, under contract with the National Aeronautics and Space Administration, and the SIMBAD database, operated at CDS, Strasbourg, France.

This work has utilized the following software: Jupyter \citep{Kluyver2016}, Astropy \citep{Astropy2013, Price-Whelan2018}, Pandas \citep{McKinney2010}, NumPy \citep{VanDerWalt2011, Harris2020}, SciPy \citep{Virtanen2020}, Seaborn \citep{Waskom2017}, Matplotlib \citep{Hunter2007}, \& GitHub \citep{github}.


\bibliography{main.bbl}


\appendix
\label{appendix}

Table \ref{tab:OSCSample} presents the full list of OSC objects (accessed on 2021 May 25) that occurred within the PHANGS--ALMA footprint and have a listed discovery date. We include the SN type, host galaxy, r.a. and dec. of the SN, and we include the reference paper for the SN type or the discovery announcement if a reference paper with typing is not available. Note that for two of the OSC objects (PTSS-19clju and PSN+204.332083-29.896833) we are unable to find a reference, and those are not included in our final, working sample. Finally, we note whether the object is included in our working SN sample.

Table \ref{tab:intensities} presents the CO~(2-1) line-integrated intensity measurements at the locations of our working SN sample in units of K~km~s$^{-1}$ at 60, 90, 120, and 150~pc resolution. 

\onecolumngrid

\renewcommand{\thefigure}{A\arabic{figure}}
\setcounter{figure}{0}
\renewcommand{\thetable}{A\arabic{table}}
\setcounter{table}{0}
\renewcommand{\theequation}{A\arabic{equation}}
\setcounter{equation}{0}

\clearpage

\startlongtable
\begin{deluxetable*}{lcccccc}
\tablewidth{0pt}
\tabletypesize{\small}
\tablecaption{Open Supernova Catalogue Objects in PHANGS--ALMA Footprint
\label{tab:OSCSample}}
\tablehead{
\colhead{Supernova} &
\colhead{Type} & 
\colhead{Galaxy} & 
\colhead{Ra} & 
\colhead{Dec} & 
\colhead{Reference Paper} &
\colhead{In Sample}} 
\startdata
AT2020juh & Candidate & circinus & 213.3379 & -65.3413 & \cite{Valenti2020} & - \\
SN1996cr & II & circinus & 213.2918 & -65.3457 & \cite{Maza1999} & $\checkmark$ \\
AT2019npi & Candidate & ngc0247 & 11.7453 & -20.7083 & \cite{Andreoni2019a} & - \\
AT2019npd & Candidate & ngc0253 & 11.7363 & -25.3768 & \cite{Andreoni2019b} & - \\
AT2020hol & Candidate & ngc0253 & 11.8539 & -25.3571 & \cite{Chambers2019} & - \\
AT2019pck & Candidate & ngc0253 & 11.8514 & -25.2261 & \cite{Andreoni2019c} & - \\
SN1940E & I & ngc0253 & 11.8783 & -25.2934 & \cite{Tsvetkov1993} & $\checkmark$ \\
SN2019qyl & IIn/LBV & ngc0300 & 13.7399 & -37.6444 & \cite{Andrews2019} & - \\
SN2013ej & II & ngc0628 & 24.2007 & 15.7586 & \cite{Valenti2013} & $\checkmark$ \\
SN2003gd & II & ngc0628 & 24.1777 & 15.739 & \cite{Evans2003} & - \\
SN2019krl & IIn/LBV & ngc0628 & 24.2068 & 15.7795 & \cite{Ho2019} & - \\
SN2018ivc & II & ngc1068 & 40.672 & -0.0088 & \cite{Ochner2018} & $\checkmark$ \\
SN1995V & II & ngc1087 & 41.6115 & -0.4988 & \cite{Evans1995} & $\checkmark$ \\
SN1999eu & II & ngc1097 & 41.5866 & -30.3184 & \cite{Nakano1999} & $\checkmark$ \\
SN1992bd & II & ngc1097 & 41.5792 & -30.2756 & \cite{Smith1992} & $\checkmark$ \\
SN2012fr & Ia & ngc1365 & 53.4006 & -36.1268 & \cite{Klotz2012} & $\checkmark$ \\
SN2001du & II & ngc1365 & 53.3713 & -36.1421 & \cite{Smartt2001} & $\checkmark$ \\
SN1983V & Ic & ngc1365 & 53.3819 & -36.1486 & \cite{Wheeler1987} & $\checkmark$ \\
SN1985P & II & ngc1433 & 55.5264 & -47.21 & \cite{Kirshner1985} & $\checkmark$ \\
SN1935C & unclassified & ngc1511 & 59.9373 & -67.6374 & \cite{vandenBergh1988} & $\checkmark$ \\
SN2009ib & II & ngc1559 & 64.4167 & -62.7776 & \cite{Pignata2009} & $\checkmark$ \\
SN2005df & Ia & ngc1559 & 64.4077 & -62.7693 & \cite{Wang2005} & $\checkmark$ \\
SN1986L & II & ngc1559 & 64.374 & -62.7846 & \cite{LloydEvans1986} & $\checkmark$ \\
SN1984J & II & ngc1559 & 64.3981 & -62.7855 & \cite{Buta1984} & $\checkmark$ \\
ASASSN-14ha & II & ngc1566 & 65.0059 & -54.9381 & \cite{Arcavi2014} & $\checkmark$ \\
SN2010el & Ia & ngc1566 & 64.9951 & -54.944 & \cite{Bessell2010} & $\checkmark$ \\
SN1999em & II & ngc1637 & 70.3629 & -2.8626 & \cite{Jha1999} & $\checkmark$ \\
SN2017gax & Ib/c & ngc1672 & 71.4561 & -59.2451 & \cite{Jha2017} & $\checkmark$ \\
SN1993Z & Ia & ngc2775 & 137.5816 & 7.0278 & \cite{Treffers1993} & $\checkmark$ \\
SN2003jg & Ic & ngc2997 & 146.4086 & -31.1891 & \cite{Howell2003} & $\checkmark$ \\
SN2008eh & unclassified & ngc2997 & 146.4507 & -31.1791 & \cite{Monard2008b} & $\checkmark$ \\
SN2012A & II & ngc3239 & 156.2808 & 17.1541 & \cite{Cao2012} & $\checkmark$ \\
SN1989B & Ia & ngc3627 & 170.058 & 13.0053 & \cite{Marvin1989} & $\checkmark$ \\
SN1997bs & IIn & ngc3627 & 170.0593 & 12.9721 & \cite{Treffers1997} & - \\
SN2016cok & II & ngc3627 & 170.0797 & 12.9825 & \cite{Zhang2016} & $\checkmark$ \\
SN2009hd & II & ngc3627 & 170.0707 & 12.9796 & \cite{Kasliwal2009} & $\checkmark$ \\
SN1973R & II & ngc3627 & 170.0481 & 12.9977 & \cite{Ciatti1977} & $\checkmark$ \\
AT2020cwh & Candidate & ngc3627 & 170.0833 & 12.9903 & \cite{Stevenson2020} & - \\
SN2014L & Ic & ngc4254 & 184.7029 & 14.4121 & \cite{Yamaoka2014} & $\checkmark$ \\
SN1986I & II & ngc4254 & 184.7169 & 14.4123 & \cite{Pennypacker1986} & $\checkmark$ \\
SN1972Q & II & ngc4254 & 184.7107 & 14.4443 & \cite{Barbon1973} & $\checkmark$ \\
SN1967H & II & ngc4254 & 184.7184 & 14.414 & \cite{Fairall1972} & $\checkmark$ \\
SN2014dt & Ia & ngc4303 & 185.4899 & 4.4718 & \cite{Ochner2014} & $\checkmark$ \\
SN1999gn & II & ngc4303 & 185.4876 & 4.4627 & \cite{Ayani1999} & $\checkmark$ \\
SN2006ov & II & ngc4303 & 185.4804 & 4.488 & \cite{Blondin2006} & $\checkmark$ \\
SN2020jfo & II & ngc4303 & 185.4602 & 4.4817 & \cite{Perley2020} & $\checkmark$ \\
SN1926A & II & ngc4303 & 185.4754 & 4.4934 & \cite{Tsvetkov1993} & $\checkmark$ \\
SN1961I & II & ngc4303 & 185.5018 & 4.4704 & \cite{Porter1993} & $\checkmark$ \\
SN1964F & II & ngc4303 & 185.4698 & 4.4738 & \cite{Tsvetkov1993} & $\checkmark$ \\
SN2006X & Ia & ngc4321 & 185.7249 & 15.809 & \cite{Quimby2006} & $\checkmark$ \\
SN1979C & II & ngc4321 & 185.7442 & 15.7978 & \cite{Carney1980} & $\checkmark$ \\
SN2020oi & Ic & ngc4321 & 185.7289 & 15.8236 & \cite{Siebert2020} & $\checkmark$ \\
SN2019ehk & Ib & ngc4321 & 185.7339 & 15.8261 & \cite{De2021} & $\checkmark$ \\
SN1901B & I & ngc4321 & 185.6971 & 15.8238 & \cite{Tsvetkov1993} & $\checkmark$ \\
PTSS-19clju & unclassified & ngc4321 & 185.7341 & 15.8261 & ? & - \\
SN1959E & I & ngc4321 & 185.7454 & 15.817 & \cite{Porter1993} & $\checkmark$ \\
SN2012cg & Ia & ngc4424 & 186.8035 & 9.4203 & \cite{Kandrashoff2012} & $\checkmark$ \\
SN2020nvb & Ia & ngc4457 & 187.245 & 3.5728 & \cite{Waagen2020} & $\checkmark$ \\
SN1960F & Ia & ngc4496a & 187.9252 & 3.9466 & \cite{Porter1993} & $\checkmark$ \\
SN1988M & II & ngc4496a & 187.9206 & 3.9228 & \cite{Filippenko1988} & $\checkmark$ \\
SN1981B & Ia & ngc4536 & 188.6231 & 2.1998 & \cite{Busko1981} & $\checkmark$ \\
SN1988A & II & ngc4579 & 189.4315 & 11.8054 & \cite{Kosai1988} & $\checkmark$ \\
SN1989M & Ia & ngc4579 & 189.4196 & 11.8237 & \cite{Kharadze1989} & $\checkmark$ \\
SN2005af & II & ngc4945 & 196.1836 & -49.5666 & \cite{Filippenko2005} & - \\
SN2011ja & II & ngc4945 & 196.2963 & -49.5242 & \cite{Monard2011} & $\checkmark$ \\
SN1986G & Ia & ngc5128 & 201.4021 & -43.0317 & \cite{Wamsteker1986} & $\checkmark$ \\
SN2016adj & IIb & ngc5128 & 201.3505 & -43.016 & \cite{Thomas2016} & $\checkmark$ \\
AT2020nqq & LRN & ngc5128 & 201.4351 & -43.0036 & \cite{Tonry2020} & - \\
SN1983N & Ib & ngc5236 & 204.2135 & -29.9006 & \cite{Gaskell1986} & $\checkmark$ \\
SN1968L & II & ngc5236 & 204.252 & -29.8665 & \cite{Rosa1988} & $\checkmark$ \\
SN1923A & II & ngc5236 & 204.2883 & -29.8389 & \cite{Rosa1988} & $\checkmark$ \\
PSN+204.332083-29.896833 & candidate & ngc5236 & 204.3321 & -29.8968 & ? & - \\
AT2018eoq & candidate & ngc5236 & 204.2501 & -29.8646 & \cite{Kendurkar2018} & - \\
SN1950B & unclassified & ngc5236 & 204.2203 & -29.8655 & \cite{Rosa1988} & $\checkmark$ \\
SN1957D & unclassified & ngc5236 & 204.2647 & -29.828 & \cite{Rosa1988} & $\checkmark$ \\
SN2007it & II & ngc5530 & 214.6067 & -43.3817 & \cite{Contreras2007} & $\checkmark$ \\
SN2005at & Ic & ngc6744 & 287.4733 & -63.8231 & \cite{Schmidt2005} & $\checkmark$ \\
SN2008bk & II & ngc7793 & 359.4601 & -32.556 & \cite{Morrell2008} & $\checkmark$ \\
\enddata
\end{deluxetable*}

\clearpage
\startlongtable
\begin{deluxetable*}{lccc|c|c|cc}
\tablewidth{0pt}
\tablecaption{CO 2-1 Intensity Measurements}
\label{tab:intensities}
\tablehead{
\colhead{Supernova} &
\colhead{Type} & 
\colhead{Galaxy} & 
\multicolumn{4}{c}{CO~(2-1) Line-Integrated Intensity [K~km~s$^{-1}$]}\\
\hline
   &&&{60pc} &{90pc} & {120pc}&{150pc}}
\startdata 
SN1996cr & II & circinus & nan & nan & nan & 186.85 $\pm$ 2.14 \\
SN1940E & I & ngc0253 & nan & nan & nan & 219.3 $\pm$ 1.47 \\
SN2019qyl & II & ngc0300 & nan & -0.63 $\pm$ 0.5 & -0.34 $\pm$ 0.25 & -0.1 $\pm$ 0.16 \\
SN2013ej & II & ngc0628 & 4.94 $\pm$ 1.73 & 2.73 $\pm$ 1.26 & 1.81 $\pm$ 1.02 & 1.34 $\pm$ 0.86 \\
SN2019krl & II & ngc0628 & 0.75 $\pm$ 1.1 & 0.96 $\pm$ 0.75 & 0.91 $\pm$ 0.58 & 0.76 $\pm$ 0.47 \\
SN2018ivc & II & ngc1068 & nan & nan & nan & nan \\
SN1995V & II & ngc1087 & nan & nan & 15.47 $\pm$ 1.14 & 12.75 $\pm$ 0.91 \\
SN1992bd & II & ngc1097 & nan & nan & 282.1 $\pm$ 2.05 & 260.37 $\pm$ 1.49 \\
SN1999eu & II & ngc1097 & nan & nan & 0.0 $\pm$ 0.75 & 0.0 $\pm$ 0.53 \\
SN1983V & Ic & ngc1365 & nan & nan & 5.46 $\pm$ 1.14 & 5.27 $\pm$ 0.98 \\
SN2001du & II & ngc1365 & nan & nan & 3.07 $\pm$ 1.34 & 2.8 $\pm$ 1.14 \\
SN2012fr & Ia & ngc1365 & nan & nan & -2.88 $\pm$ 1.1 & -2.53 $\pm$ 0.92 \\
SN1985P & II & ngc1433 & nan & 0.26 $\pm$ 0.59 & 0.27 $\pm$ 0.47 & 0.24 $\pm$ 0.39 \\
SN1935C & unclassified & ngc1511 & nan & nan & 2.22 $\pm$ 0.65 & 1.64 $\pm$ 0.52 \\
SN1984J & II & ngc1559 & nan & nan & 14.34 $\pm$ 1.56 & 13.67 $\pm$ 1.18 \\
SN1986L & II & ngc1559 & nan & nan & 2.43 $\pm$ 1.21 & 2.29 $\pm$ 0.92 \\
SN2005df & Ia & ngc1559 & nan & nan & 0.0 $\pm$ 0.43 & 0.0 $\pm$ 0.33 \\
SN2009ib & II & ngc1559 & nan & nan & 3.03 $\pm$ 1.25 & 3.68 $\pm$ 0.93 \\
ASASSN-14ha & II & ngc1566 & nan & nan & 65.95 $\pm$ 1.67 & 61.93 $\pm$ 1.36 \\
SN2010el & Ia & ngc1566 & nan & nan & 14.74 $\pm$ 1.33 & 14.9 $\pm$ 1.09 \\
SN1999em & II & ngc1637 & nan & 2.99 $\pm$ 0.41 & 2.82 $\pm$ 0.29 & 2.78 $\pm$ 0.24 \\
SN2017gax & Ib/c & ngc1672 & nan & nan & nan & nan \\
SN1993Z & Ia & ngc2775 & nan & nan & -0.38 $\pm$ 1.61 & 0.15 $\pm$ 1.29 \\
SN2003jg & Ic & ngc2997 & nan & nan & 106.31 $\pm$ 1.03 & 101.07 $\pm$ 0.79 \\
SN2008eh & unclassified & ngc2997 & nan & nan & 14.22 $\pm$ 0.58 & 13.2 $\pm$ 0.47 \\
SN2012A & II & ngc3239 & nan & 0.0 $\pm$ 0.64 & 0.0 $\pm$ 0.56 & 0.0 $\pm$ 0.51 \\
SN1973R & II & ngc3627 & nan & 10.67 $\pm$ 1.21 & 11.45 $\pm$ 0.82 & 11.58 $\pm$ 0.63 \\
SN1989B & Ia & ngc3627 & nan & 10.17 $\pm$ 1.51 & 18.17 $\pm$ 1.05 & 28.7 $\pm$ 0.81 \\
SN1997bs & II & ngc3627 & nan & 3.39 $\pm$ 1.35 & 4.76 $\pm$ 1.0 & 5.64 $\pm$ 0.8 \\
SN2009hd & II & ngc3627 & nan & 68.4 $\pm$ 1.86 & 96.1 $\pm$ 1.32 & 114.62 $\pm$ 1.06 \\
SN2016cok & II & ngc3627 & nan & 5.64 $\pm$ 1.21 & 5.62 $\pm$ 0.87 & 5.22 $\pm$ 0.69 \\
SN1967H & II & ngc4254 & nan & nan & 13.98 $\pm$ 0.83 & 14.16 $\pm$ 0.59 \\
SN1972Q & II & ngc4254 & nan & nan & 2.59 $\pm$ 0.63 & 2.23 $\pm$ 0.44 \\
SN1986I & II & ngc4254 & nan & nan & 8.62 $\pm$ 0.9 & 10.53 $\pm$ 0.66 \\
SN2014L & Ic & ngc4254 & nan & nan & 43.08 $\pm$ 1.01 & 45.1 $\pm$ 0.73 \\
SN1926A & II & ngc4303 & nan & nan & nan & -0.73 $\pm$ 0.84 \\
SN1961I & II & ngc4303 & nan & nan & nan & 1.76 $\pm$ 1.95 \\
SN1964F & II & ngc4303 & nan & nan & nan & 2.98 $\pm$ 0.99 \\
SN1999gn & II & ngc4303 & nan & nan & nan & 6.38 $\pm$ 0.9 \\
SN2006ov & II & ngc4303 & nan & nan & nan & -0.63 $\pm$ 1.25 \\
SN2014dt & Ia & ngc4303 & nan & nan & nan & 0.83 $\pm$ 0.98 \\
SN2020jfo & II & ngc4303 & nan & nan & nan & 1.05 $\pm$ 0.99 \\
SN1901B & I & ngc4321 & nan & nan & 5.66 $\pm$ 1.03 & 5.48 $\pm$ 0.83 \\
SN1959E & I & ngc4321 & nan & nan & 18.03 $\pm$ 1.24 & 17.46 $\pm$ 0.99 \\
SN1979C & II & ngc4321 & nan & nan & 1.86 $\pm$ 0.79 & 1.53 $\pm$ 0.65 \\
SN2006X & Ia & ngc4321 & nan & nan & 0.13 $\pm$ 1.12 & 0.66 $\pm$ 0.9 \\
SN2019ehk & Ib & ngc4321 & nan & nan & 1.51 $\pm$ 1.64 & 0.9 $\pm$ 1.28 \\
SN2020oi & Ic & ngc4321 & nan & nan & 14.0 $\pm$ 2.05 & 18.6 $\pm$ 1.65 \\
SN2012cg & Ia & ngc4424 & nan & -0.52 $\pm$ 0.98 & -0.7 $\pm$ 0.7 & -0.59 $\pm$ 0.58 \\
SN2020nvb & Ia & ngc4457 & nan & 14.36 $\pm$ 1.95 & 13.19 $\pm$ 1.47 & 12.4 $\pm$ 1.22 \\
SN1960F & Ia & ngc4496a & nan & -0.15 $\pm$ 0.82 & 0.0 $\pm$ 0.64 & 0.01 $\pm$ 0.56 \\
SN1988M & II & ngc4496a & nan & 0.0 $\pm$ 0.52 & 0.0 $\pm$ 0.4 & 0.0 $\pm$ 0.35 \\
SN1981B & Ia & ngc4536 & nan & nan & 0.04 $\pm$ 0.34 & 0.2 $\pm$ 0.26 \\
SN1988A & II & ngc4579 & nan & nan & nan & nan \\
SN1989M & Ia & ngc4579 & nan & nan & nan & nan \\
SN2011ja & II & ngc4945 & nan & nan & 48.65 $\pm$ 0.72 & 48.28 $\pm$ 0.6 \\
SN1986G & Ia & ngc5128 & nan & nan & nan & 9.31 $\pm$ 0.98 \\
SN2016adj & IIb & ngc5128 & nan & nan & nan & 67.57 $\pm$ 1.43 \\
SN1923A & II & ngc5236 & 2.98 $\pm$ 0.59 & 2.65 $\pm$ 0.46 & 2.57 $\pm$ 0.36 & 2.6 $\pm$ 0.29 \\
SN1950B & unclassified & ngc5236 & 61.97 $\pm$ 0.65 & 66.93 $\pm$ 0.46 & 65.22 $\pm$ 0.36 & 60.49 $\pm$ 0.29 \\
SN1957D & unclassified & ngc5236 & 38.48 $\pm$ 0.53 & 31.96 $\pm$ 0.39 & 27.75 $\pm$ 0.3 & 24.64 $\pm$ 0.24 \\
SN1968L & II & ngc5236 & 164.3 $\pm$ 0.96 & 181.97 $\pm$ 0.62 & 202.79 $\pm$ 0.45 & 222.91 $\pm$ 0.36 \\
SN1983N & Ib & ngc5236 & 2.45 $\pm$ 0.62 & 2.1 $\pm$ 0.48 & 1.77 $\pm$ 0.37 & 1.52 $\pm$ 0.29 \\
SN2007it & II & ngc5530 & nan & 2.92 $\pm$ 1.04 & 2.72 $\pm$ 0.83 & 2.7 $\pm$ 0.71 \\
SN2005at & Ic & ngc6744 & 7.0 $\pm$ 1.23 & 6.57 $\pm$ 0.8 & 6.09 $\pm$ 0.61 & 5.61 $\pm$ 0.49 \\
SN2008bk & II & ngc7793 & nan & nan & nan & -0.01 $\pm$ 0.17
\enddata
\end{deluxetable*}
\section{}
\end{document}